\documentclass[twocolumn]{aastex62}
\usepackage{natbib}
\graphicspath{{./}{../pictures/}}

\received{}
\revised{}
\accepted{}
\submitjournal{ApJL}

\shorttitle{Extreme tidal disruption events}
\shortauthors{Ryu et al.}

\usepackage{graphicx}	
\usepackage{amsmath}	
\usepackage{amssymb}	
\usepackage{cases}
\usepackage{array,multirow}
\usepackage{upgreek}
\usepackage{textcomp}

\usepackage{hyperref}
\usepackage[referable]{threeparttablex}
\usepackage{booktabs, dcolumn}
\makeatletter
\newcommand*{\rom}[1]{\expandafter\@slowromancap\romannumeral #1@}
\makeatother

\newcommand{\beq}{\begin{equation}}
\newcommand{\eeq}{\end{equation}}
\newcommand{\simlt}{\mathrel{\hbox{\rlap{\hbox{\lower4pt\hbox{$\sim$}}}\hbox{$<$}}}}
\newcommand{\simgt}{\mathrel{\hbox{\rlap{\hbox{\lower4pt\hbox{$\sim$}}}\hbox{$>$}}}}

\newcommand{\Msol}{\;\mathrm{M}_{\odot}}

\def\apjl{ApJL}
\def\apj{ApJ}
\def\mnras{M.N.R.A.S.}
\def\aap{A\&A}
\def\nat{Nat.}

\def\araa{Ann. Rev. A\&A}

\def\apjs{ApJ Supp.}

\usepackage[normalem]{ulem}
\usepackage{booktabs,tikz}

\newcommand{\harm}{{\sc Harm3d}}   
\newcommand{\mesa}{{\small MESA}}

\begin{document}

\title{Extremely Relativistic Tidal Disruption Events}

\correspondingauthor{Taeho Ryu}
\email{tryu@mpa-garching.mpg.de}

\author[0000-0002-0786-7307]{Taeho Ryu}
\affil{The Max Planck Institute for Astrophysics, Karl-Schwarzschild-Str. 1, Garching, 85748, Germany}
\affiliation{Physics and Astronomy Department, Johns Hopkins University, Baltimore, MD 21218, USA}

\author{Julian Krolik}
\affiliation{Physics and Astronomy Department, Johns Hopkins University, Baltimore, MD 21218, USA}
\author{Tsvi Piran}
\affiliation{Racah Institute of Physics, Hebrew University, Jerusalem 91904, Israel}

\begin{abstract}

Extreme tidal disruption events (eTDEs), which occur when a star passes very close to a supermassive black hole, may provide a way to observe a long-sought general relativistic effect: orbits that wind several times around a black hole and then leave. Through general relativistic hydrodynamics simulations, we show that such eTDEs are easily distinguished from most tidal disruptions, in which stars come close, but not so close, to the black hole. Following the stellar orbit, the debris is initially distributed in a crescent, it then turns into a set of tight spirals circling the black hole, which merge into a shell expanding radially outwards. Some mass later falls back toward the black hole, while the remainder is ejected. Internal shocks within the  infalling debris power the observed emission.
The resulting light-curve rises rapidly to roughly the Eddington luminosity, maintains this level for  between a few weeks and a year (depending on both the stellar mass and the black hole mass), and then drops.  Most of its power is in thermal X-rays at a temperature $\sim (1-2)\times 10^{6}$~K ($\sim 100-200$~eV).  The debris evolution and observational features of eTDEs are qualitatively different from ordinary TDEs, making eTDEs a new type of TDE. Although eTDEs are relatively rare for lower-mass black holes, most tidal disruptions around  higher-mass black holes are extreme.  Their detection offers a view of an exotic relativistic phenomenon previously inaccessible.

\end{abstract}


\section{Introduction} \label{sec:intro}

\begin{figure*}
	\centering
	\includegraphics[width=0.24\textwidth]{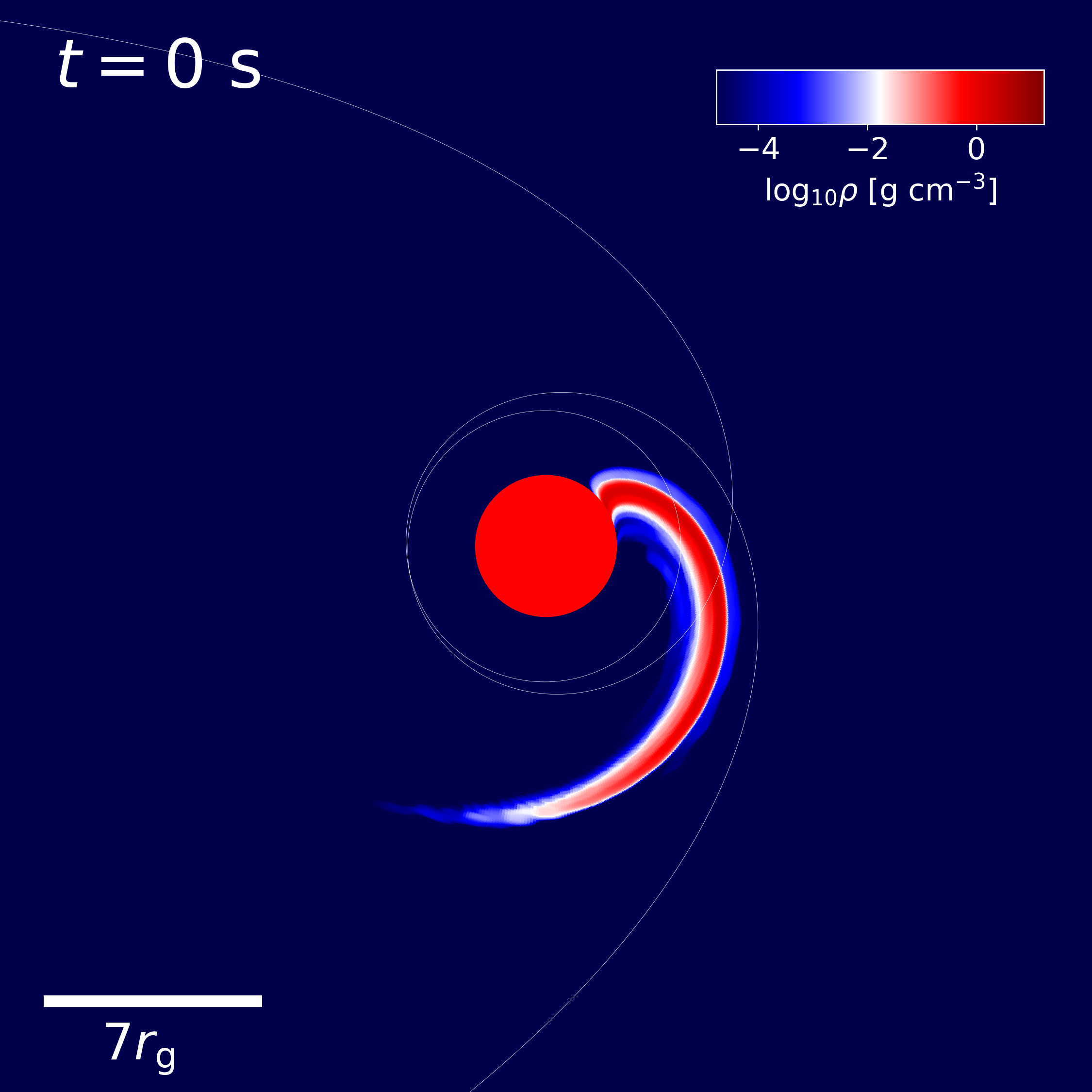}
        \includegraphics[width=0.24\textwidth]{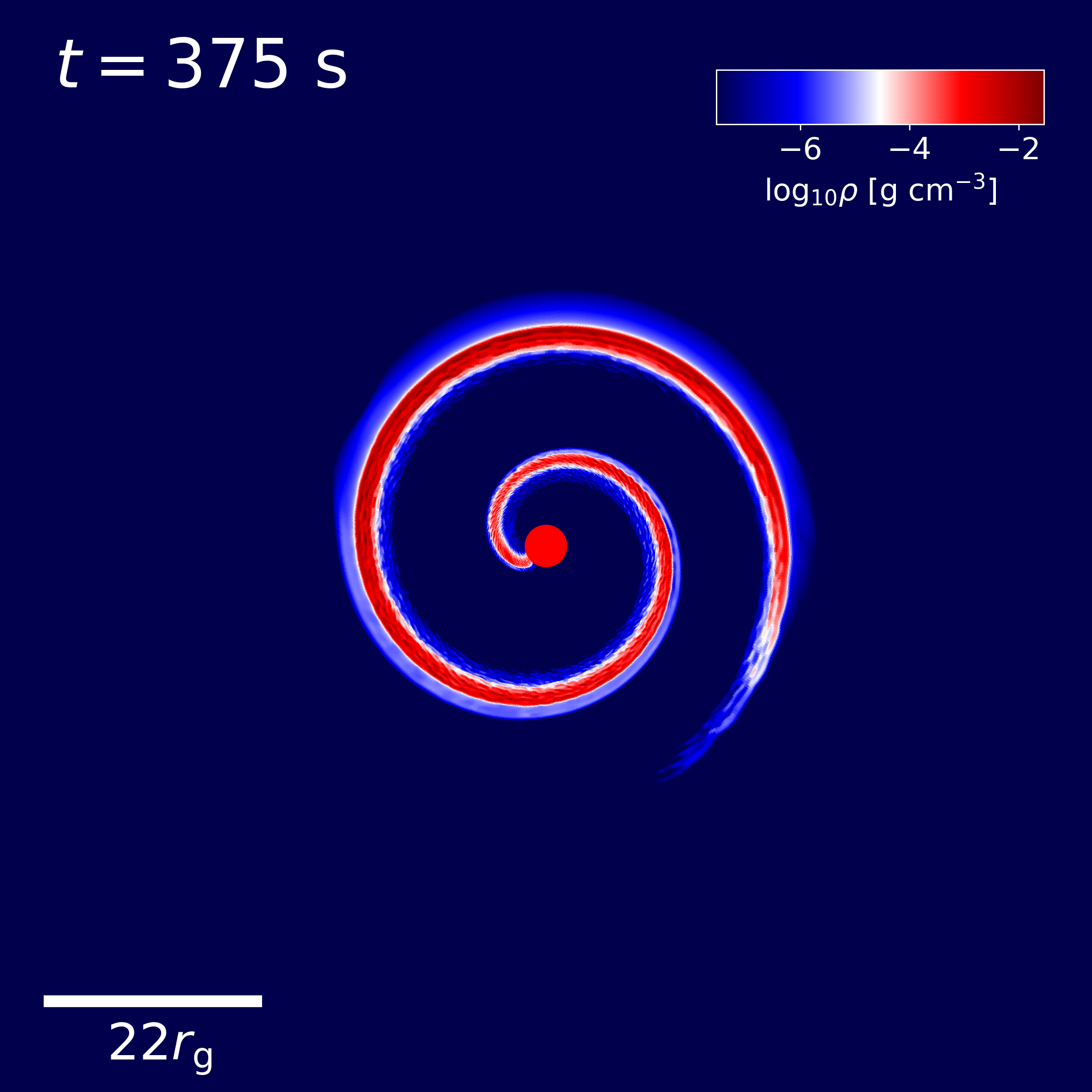}
	\includegraphics[width=0.24\textwidth]{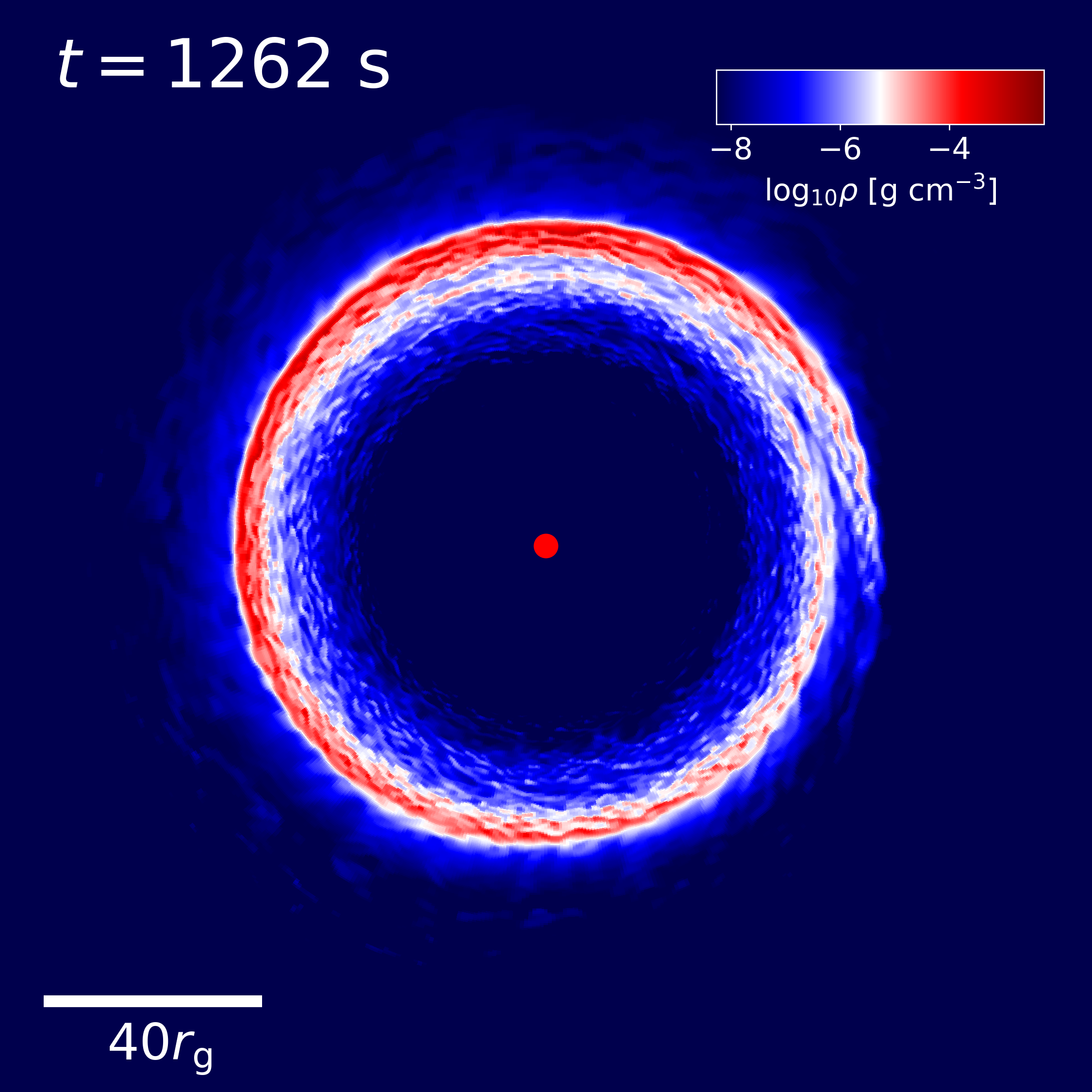}
	\includegraphics[width=0.24\textwidth]{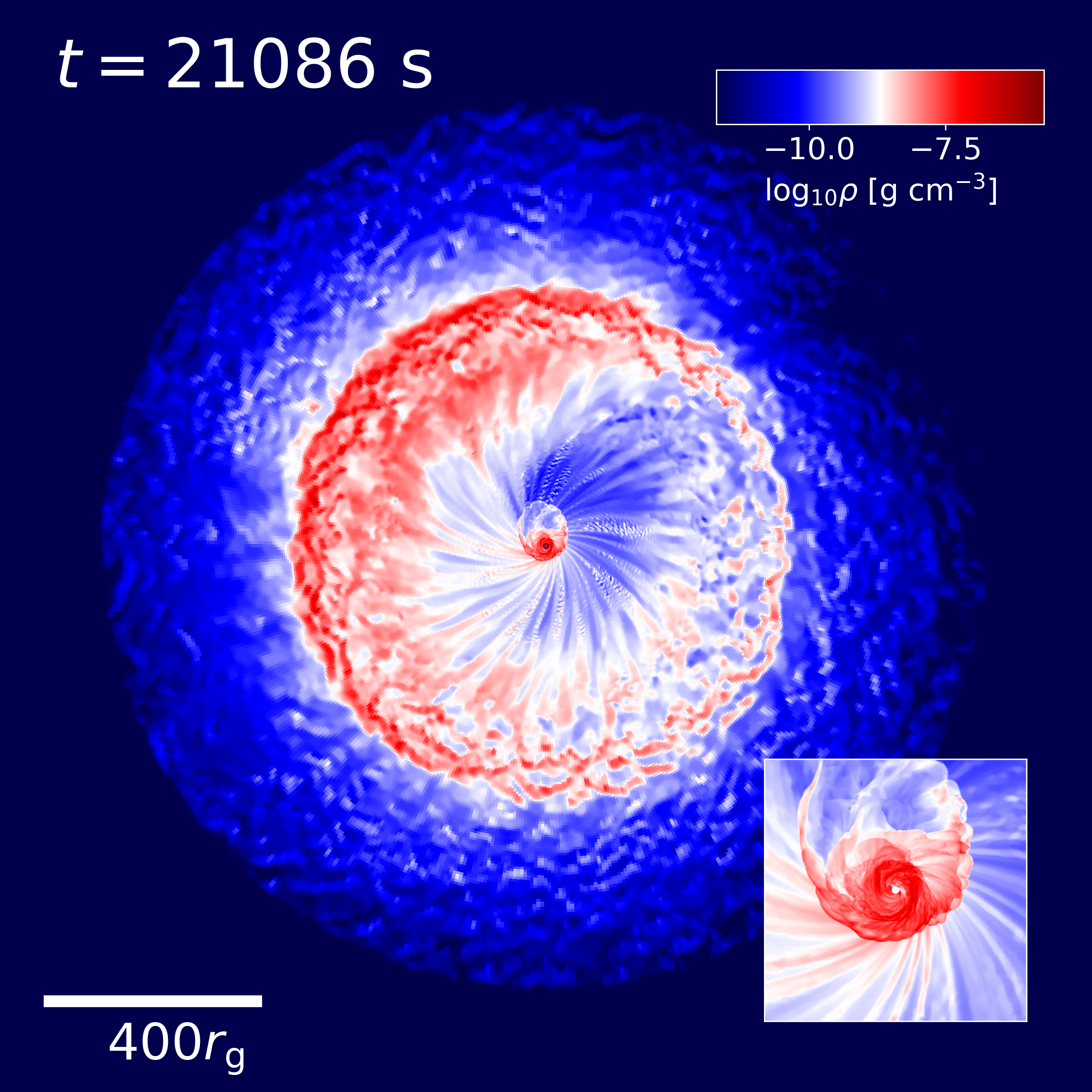}
	\includegraphics[width=0.24\textwidth]{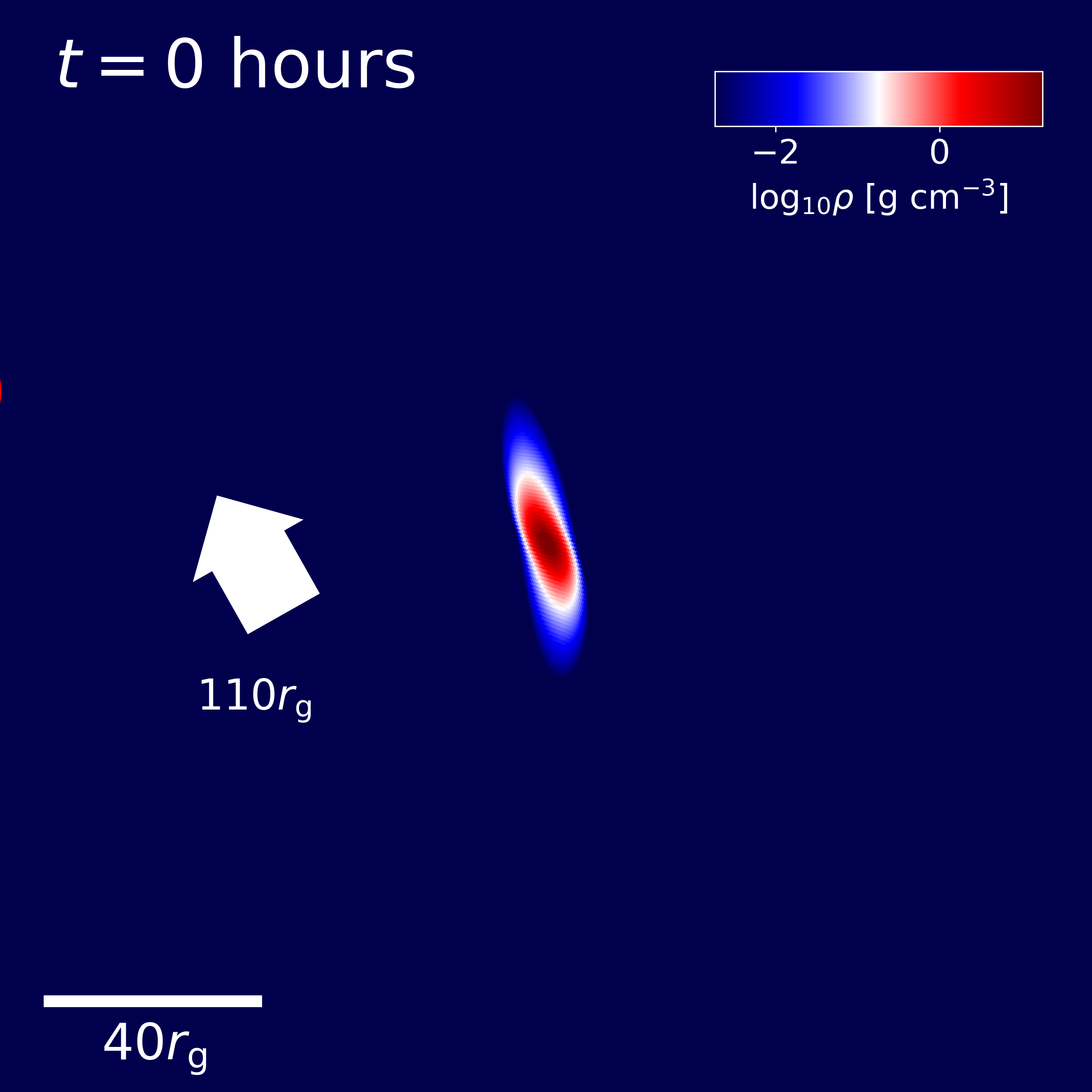}
	\includegraphics[width=0.24\textwidth]{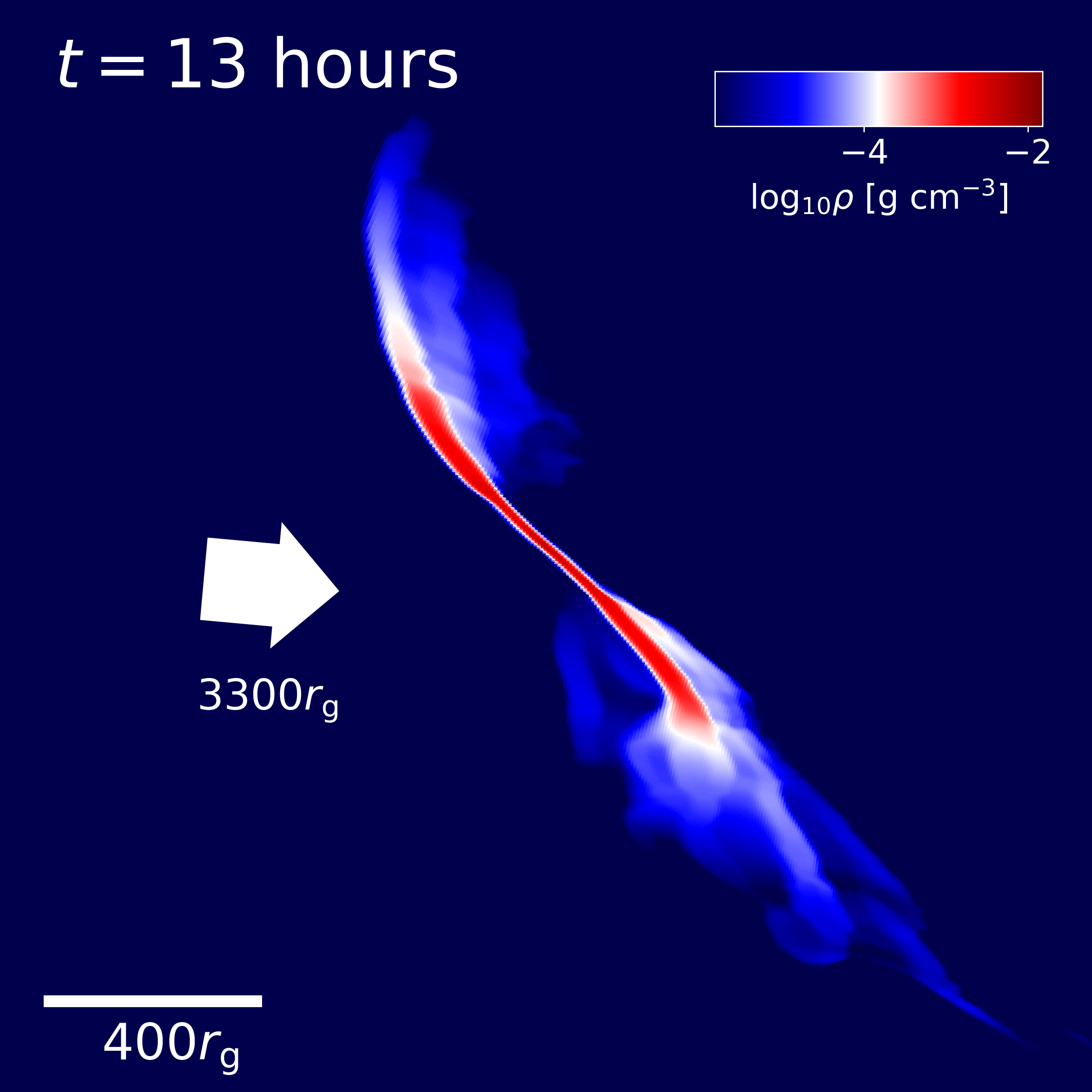}
	\includegraphics[width=0.24\textwidth]{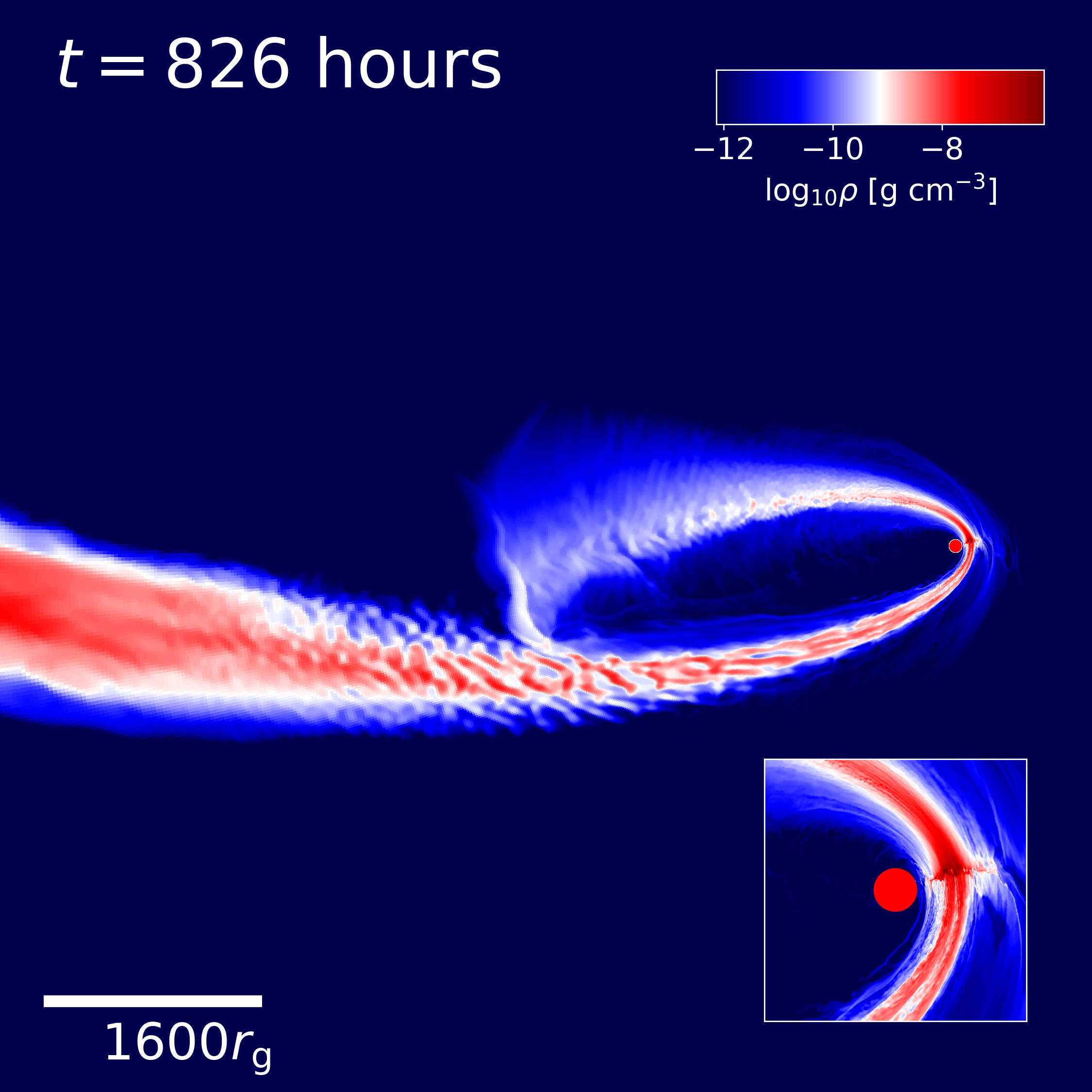}
	\includegraphics[width=0.24\textwidth]{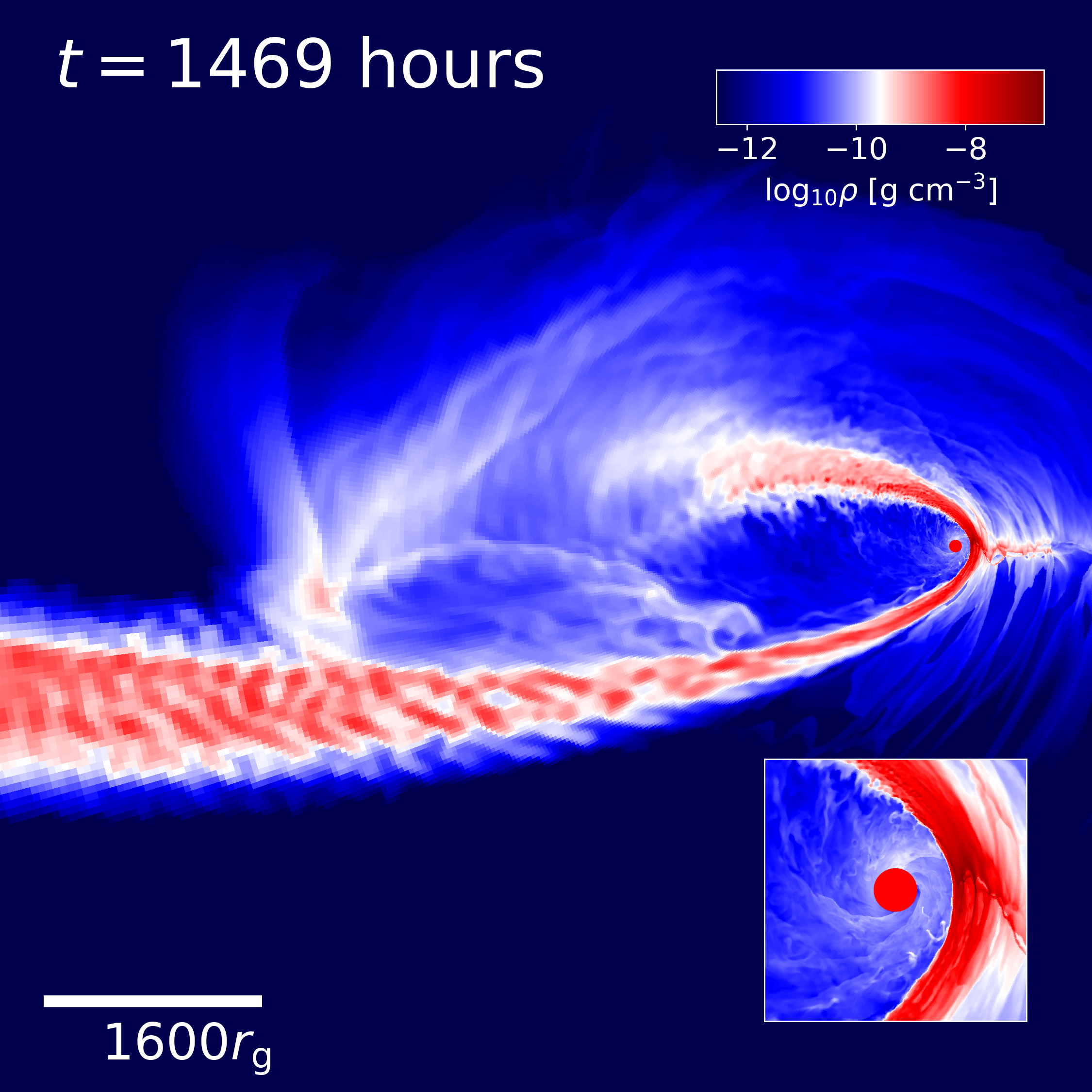}
	\caption{Illustration of the evolution and shape of debris in an extremely relativistic event  with $r_{\rm p}\sim 4.02 r_{\rm g}$ (\textit{upper} panels) and an ordinary TDE with $r_{\rm p}\sim 110 r_{\rm g}$ (\textit{lower} panels). Four phases are shown for each. For the eTDE: crescent (note that the star has already been fully disrupted at this stage and a significant fraction of its mass has been captured); spiral; ring; and ring with inflow.  For evolution of an ordinary TDE: beginning of the disruption; highly-stretched star; and two stages of the stream's return to the vicinity of the SMBH. In all cases, the colorscale represents the logarithmic density in the orbital plane. Insets show the matter near the SMBH. }
	\label{fig:debris}
\end{figure*}

Almost every galaxy harbors a supermassive black hole (SMBH) at its center \citep{KormendyHoe2013}.
Well before observational data established this fact, theoretical work 
\citep[e.g.][]{Lacy+,CarterLuminet,Rees1988} demonstrated that if a star approaches a SMBH
closer than a ``tidal radius" that is $\sim \Psi(M_\star,M_{\rm BH})  (R_\star (M_{\rm BH}/M_\star)^{1/3}$ (here $M_{\rm BH}$ and $M_\star$ are the mass of the BH and the star, respectively, $R_\star$ is the stellar radius, and $\Psi(M_\star,M_{\rm BH})$ is a correction factor of order unity \citep{Ryu1+2020}), it is disrupted by the SMBH's tidal gravity.  For $M_{\rm BH} = 10^6 M_\odot$, the critical distance for total disruption of main sequence stars is $\simeq 25r_{\rm g}$ ($r_{\rm g} \equiv GM_{\rm BH}/c^2$), nearly independent of $M_\star$ \citep{Ryu1+2020}.
In ordinary TDEs, those in which the star's pericenter $r_{\rm p}$ is not far inside the critical radius, the star follows an essentially parabolic orbit as it approaches the SMBH. After the disruption, the debris forms an elongated structure. Half the matter is unbound and rushes away, while the other half is placed on highly-eccentric ($1 - e \sim 2 (M_{\rm BH}/M_\star)^{-1/3}$) orbits. (see the lower panels of Fig.~\ref{fig:debris}). Near their apocenters, the orbits of different streams of bound matter intersect, dissipating energy with a rest-mass efficiency $\sim 10^{-4} - 10^{-3}$.
The result of these interactions is an irregular, crudely elliptical accretion flow \citep[e.g.,][]{Shiokawa+2015,Piran2015,Svirski2017,SteinbergStone2022}.

In the last $\simeq 15$~yr, roughly 100 such events have been observed \citep{Gezari2021}, generally producing an optical/UV luminosity similar to what the stream-intersections would yield \citep{Piran2015}.
The luminosity of such a flare grows on the timescale of the orbital period of the most-bound debris,
$t_0 \sim 1 ~{\rm month} ~ (M_{\rm BH}/10^6M_{\odot})^{1/2} (M_\star/M_\odot)^{-1}(R_\star/R_{\odot})^{3/2} \Xi(M_\star,M_{\rm BH})^{-3/2}$ month \citep{Rees1988}, where $\Xi$ is an order-unity correction \citep{Ryu1+2020}.
After the peak is reached, the rate at which bound mass returns to the neighborhood of the SMBH declines $\propto t^{-5/3}$ \citep{Rees1988,Phinney1989}, and many (but by no means all) observed TDE lightcurves follow this trend \citep{KomossaBade1999,Halpern+2004,Hung+2017,vanVelzen+2021}. 

\begin{figure}
	\centering
	\includegraphics[height=8.cm,angle=0]{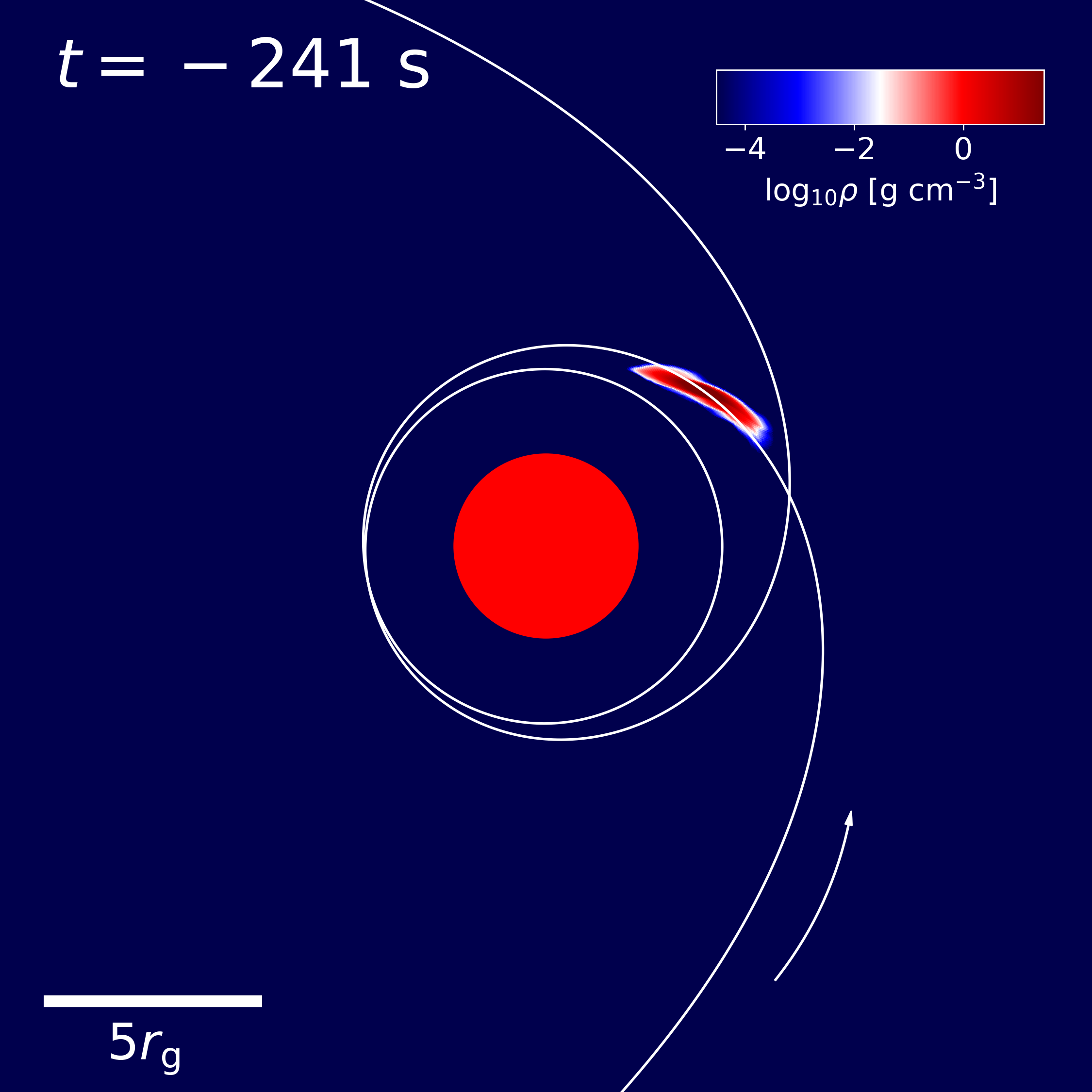}
	\caption{
    The solid white curve depicts the geodesic of an orbit with $r_{\rm p} \simeq 4.03 r_{\rm g}$ around a SMBH (red disk at the center); 
    the arrow indicates the direction of the orbit.  The color-scale shows the density distribution of stellar debris 241~s before a star whose center of mass follows this geodesic passes through pericenter. }
	\label{fig:trajectory}
\end{figure}

Remarkably, even though a SMBH causes the tidal disruption, in ordinary TDEs much of the subsequent evolution of the debris  can be explored using Newtonian dynamics. However, general relativity changes the character of orbits dramatically when their pericenter distance is $< 6 r_{\rm g}$.
When a star falls from far away with a total energy very close to its rest-mass energy and passes this close to a SMBH, rather than tracing a parabola as it would under Newtonian gravity, relativistic apsidal precession is so strong that the pericenter region wraps  all the way around the SMBH (Fig.~\ref{fig:trajectory}).  In extreme cases, the orbit can go several times  around the SMBH while keeping a distance just slightly greater than the pericenter. Only after completing these circuits can the orbital path once again extend out to large distance.  When a star follows such an orbit, the time during which it suffers extremely strong tidal gravity can be is greatly extended, an effect that, as we will show here, dramatically alters the fate of its post-disruption debris.

Several earlier works investigated the initial stage of stream evolution in such extreme disruptions. \citet{Laguna+1993} were the first to simulate eTDEs, considering a case with $r_{\rm p} = 4.7r_{\rm g}$. 
\citet{Kobayashi+2004} reconsidered the same event
focusing on the gravitational wave signature
during the strongest compression of debris at the first pericenter passage. Later, eTDEs have been simulated to examine the impact of relativity on the energy  and angular momentum distributions of the debris immediately after it leaves the star \citep[e.g.,][]{Cheng2014} and to compare the initial stage of stream evolution in non-spinning and spinning SMBHs \citep[e.g.,][]{Tejeda+2017,GaftonRosswog2019}. All these previous studies found that immediately after disruption the debris forms a crescent around the SMBH; those running a little bit longer found that the crescent becomes a spiral.
However, all stopped  when the debris was still close to the SMBH.

 Other studies considered stars on orbits with pericenters $7 r_{\rm g} \leq r_{\rm p}\leq 20r_{\rm g}$ passing by a black hole with mass $\sim 10^5 M_\odot$ \citep[e.g,][]{Evans+2015,Darbha+2019}.  In these cases, a small part of the star came close to the black hole, but the majority was too far away to reveal the effects we discuss here. In fact,
the debris in these simulations does not form a crescent;
instead it resembles the common TDE debris structure.

Here, we report on the first simulations  that follow the evolution of the debris from an eTDE long enough to  estimate  the observational signature. Our simulations, which are fully-relativistic, continue far beyond 
the longest end-point of previous work.
We find that at later times the debris undergoes multiple shape  transitions,  which ultimately lead to formation of a hot accretion flow near the SMBH (see Fig.~\ref{fig:debris}).
This inner hot flow is the main source powering the event's flare,
whose observational signature, 
both lightcurve and spectrum, are very different from those observed in  ordinary TDEs.

Our paper is organized as follows: we begin with a detailed description of the numerical methods in Sec.~\ref{sec:method}. Our results are presented in Sec.~\ref{sec:results}, and their implications are discussed in Sec.~\ref{sec:discussion}. We summarize and conclude in Sec.~\ref{sec:summary}.

\section{Numerical Methods}\label{sec:method}

We performed a series of fully relativistic hydrodynamics simulations of tidal disruptions of a realistic star on very deeply plunging zero binding-energy orbits around a $10^{6} M_{\odot}$ SMBH, using the grid-based code {\sc Harm3D}~\citep{Noble+2009}. 
 As described in \citet{Ryu2+2020}, the initial state of all the stars was taken from a stellar model for a $1M_{\odot}$ middle-aged main-sequence star evolved using the stellar evolution code {\sc MESA}~ \citep{Paxton+2011}.

The first stage of our calculations uses the \citet{Ryu2+2020} method, in which the star's dynamics are computed in a Cartesian domain that extends $5R_{\star}$ in each dimension and follows the star's center of mass along its geodesic until the star is completely disrupted.  In this approach, the star's self-gravity is calculated with the Newtonian Poisson equation in an orthonormal tetrad frame comoving with the star.  Because the metric in this frame departs from Minkowski by very small amounts within the simulation domain, the potential can be added as a perturbation to $g_{\rm tt}$. The modified tetrad-frame metric is then transformed back to the simulation coordinates. This procedure ensures that the self-gravity calculations are consistent with relativity.
Although the star becomes strongly distorted during this stage, negligible mass is lost from the box.

The second stage of the calculation begins when the tidal force completely dominates the self-gravity (at $r\lesssim 5-6r_{\rm g}$). At this point, the tidal force is more than an order of magnitude greater than the self-gravity even a single cell away from the debris' center of mass.
We therefore switch off the self-gravity,
interpolate data from the box's Cartesian grid into the spherical grid,
and  continue to follow the evolution of the debris on a spherical grid that covers the entire volume near the SMBH (for details, see Appendix~\ref{appendix:numericalgrid}). 
Self-gravity remains unimportant even in the long-term evolution of the debris because multiple shocks due to stream-stream collisions keep almost the entire debris hot.

Until the tidal force becomes dominant over the star's self-gravity, we evolve the gas using the equation of state $p=(\Gamma-1)u $ with $\Gamma = 5/3$ where $p$ is the pressure and $u$ internal energy.  When stellar self-gravity becomes negligible, we switch to an equation of state with an ``effective adiabatic index'' \citep{Shiokawa+2015} expressed as 
\begin{align}
	\Gamma =  \frac{4+5 u_{\rm gas}/u_{\rm rad}}{3(1+ u_{\rm gas}/u_{\rm rad})}.
\end{align}
This form includes radiation pressure under the assumption of  thermodynamic equilibrium.  Here, $u_{\rm gas}/u_{\rm rad}$ is the ratio of the gas internal energy density to the radiation energy density.

\section{Results}\label{sec:results}

Comparing the evolution of debris with  four different values of $r_{\rm p}=$ $4.03$, $5$, $6$ and $7r_{\rm g}$ ($L\simeq 4.0-4.5 r_{\rm g}c$), we find that extreme apsidal precession for $r_{\rm p}< 6r_{\rm g}$ (precession angle $> \uppi/2$) causes debris evolution qualitatively different from ordinary disruption events that take place at larger $r_{\rm p}$. The essential element is an orbit that stays very close to the SMBH for at least one complete circuit.
Although in the following we describe in detail the results for 
$r_{\rm p}\simeq 4.03r_{\rm g}$, debris behavior for orbits with $r_{\rm p}\lesssim 6r_{\rm g}$ is qualitatively similar.  In sharp contrast, orbits with $r_{\rm p}\simeq 7r_{\rm g}$ produce debris flows akin to ordinary TDEs.

\begin{figure}
	\centering
	\includegraphics[height=6.5cm,angle=0]{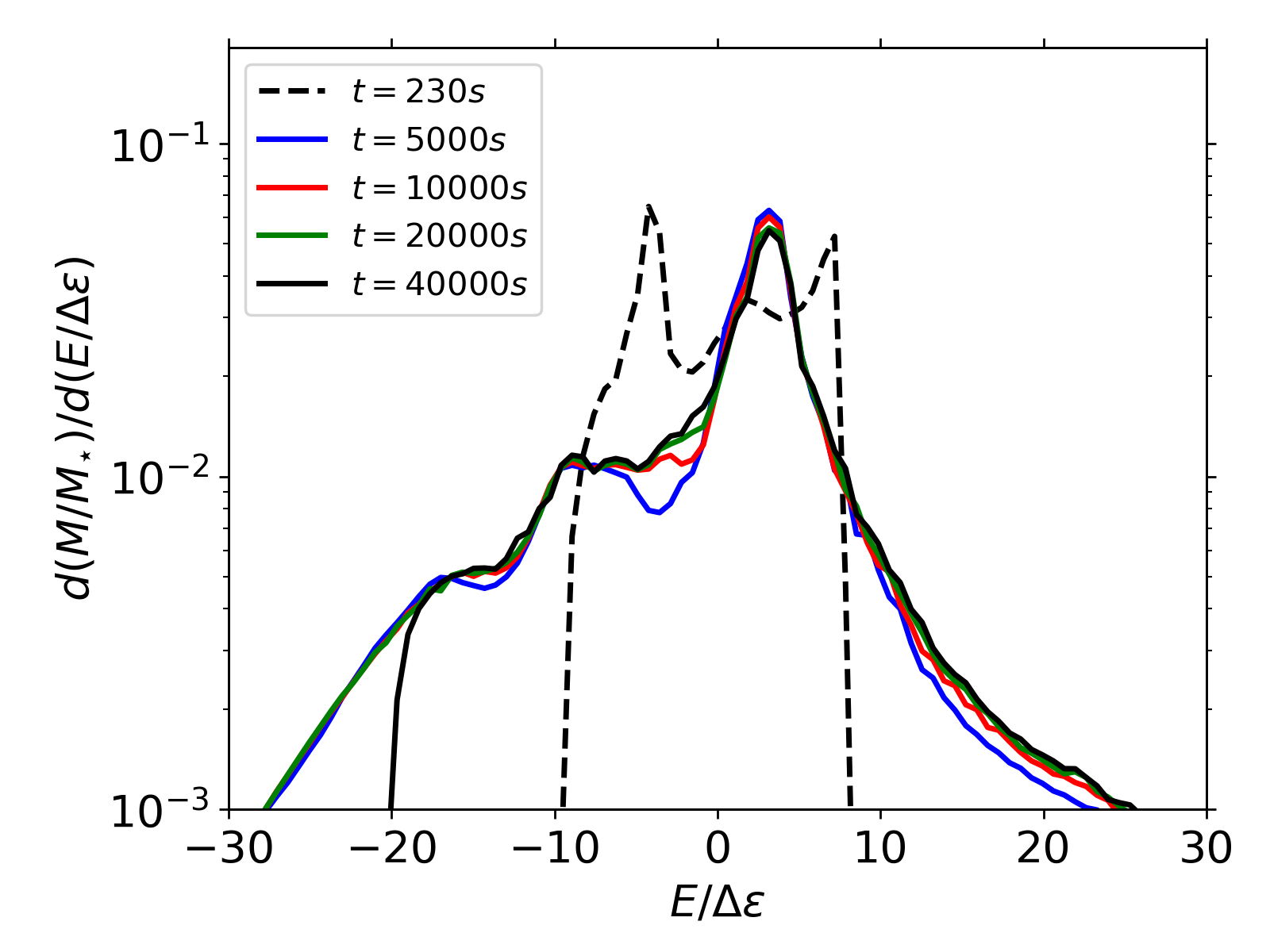}
	\includegraphics[height=6.5cm,angle=0]{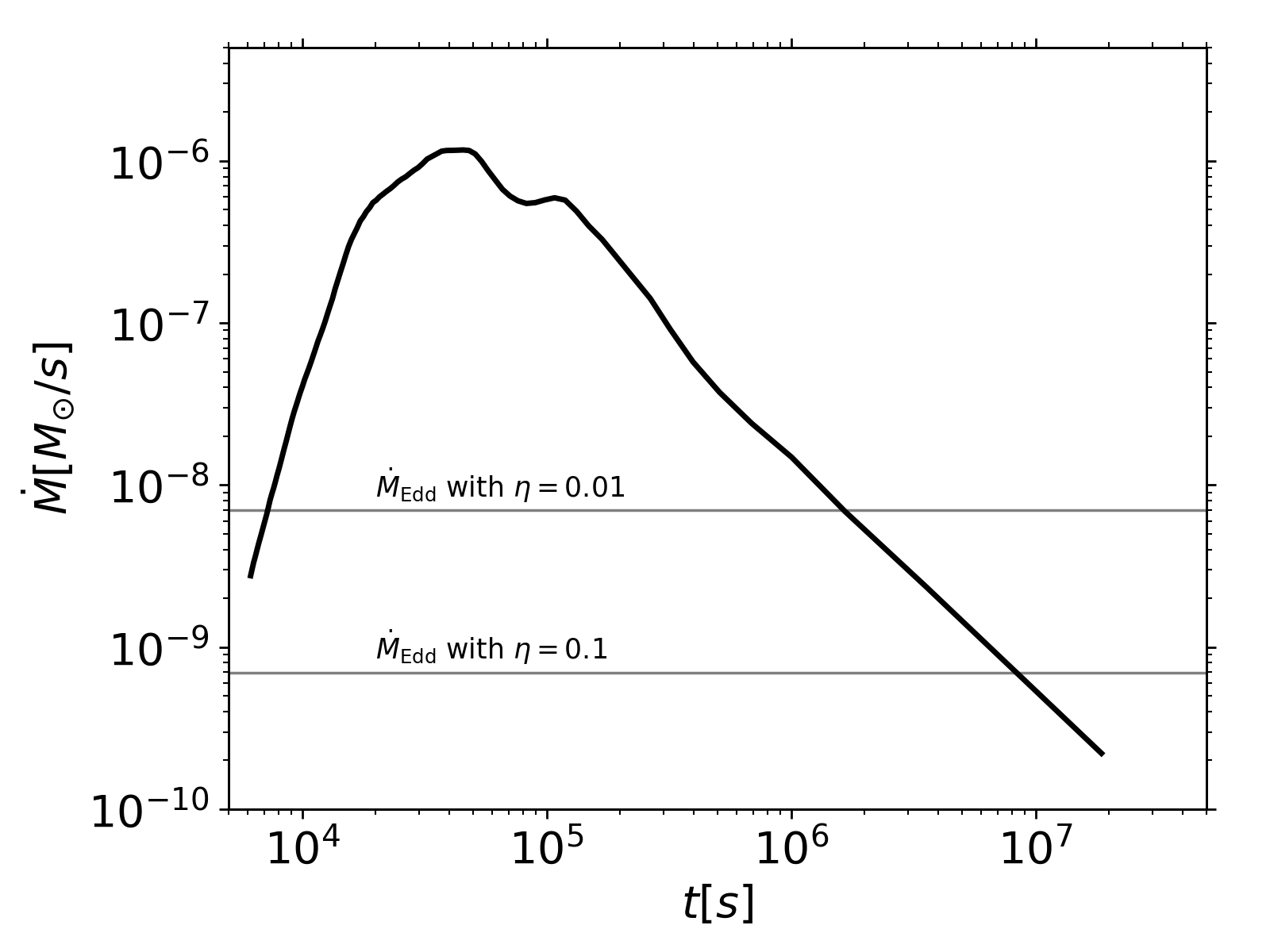}
	\caption{
	(Upper panel.)
	The orbital energy distribution of the debris from a $1 M_\odot$
	star disrupted after following an orbit with $r_{\rm p}\simeq 4.03r_{\rm g}$ around a SMBH with $M_{\rm BH} = 10^{6}M_{\odot}$.
	Energy is in units of $\Delta\epsilon$.
	The black dashed line shows the energy distribution in the gas remaining immediately after half the stellar mass plunges directly into the SMBH; the four colored curves show its evolution at later times. Here, $E=-(1+u_{\rm t})$ and $\Delta\epsilon = GM_{\rm BH}R_{\star}/r_{\rm t}^{2}$ where $r_{\rm t}=(M_{\rm BH}/M_{\star})^{1/3}R_{\star}$. The time (in seconds) is measured since the initial pericenter passage. (Bottom panel.) The thick black curve shows the mass fallback rate predicted from the energy distribution as of $t=40000$~s.
	The thin gray lines indicate the Eddington accretion rate assuming radiation efficiencies of $\eta = 0.01$ or $0.1$.
	}\label{fig:B_distribution}
\end{figure}

\subsection{Overview of dynamics}\label{subsec:overview}

Fig.~\ref{fig:trajectory} depicts the geodesic trajectory of the center of mass as well as the debris just before reaching the pericenter. As shown in this figure,  the star makes two complete trips 
around the {SMBH}, maintaining a separation $\lesssim 4.1r_{\rm g}$ for nearly the entire time.  While it does so, it continually loses mass; because of the strong apsidal precession, it spends enough time very close to the SMBH before reaching pericenter that it is wholly disrupted before the original stellar trajectory would reach the pericenter.  Roughly 3/4 of the bound mass, close to half the star's initial mass, is captured immediately, some of it even before the nominal pericenter passage. 
Meanwhile, the remainder of the debris expands away from the star. As shown in the $2^{\rm nd}$ upper panel of Fig.~\ref{fig:debris}, the result is a spiral of gas around the SMBH comprising both bound and unbound gas, but predominantly the latter. As the spiral expands further, its arms merge into a hot circular ring shown in the 3$^{\rm rd}$ upper panel; strong shocks accompany this merger.  This ring then continues to expand.  Ultimately (4th upper panel), the bound matter in the ring falls back as it reaches its orbital apocenter at $\simeq 200 r_{\rm g}$, shocking upon itself as it converges toward the SMBH.  This last stage occurs (in our fiducial simulation) at $\sim 10^4$~s after the star is disrupted.  Meanwhile, the unbound matter continues to move outward.

This behavior stands in a dramatic  contrast to that of ordinary TDEs, in which the debris forms a long, narrow stream (see the $1^{\rm st}$ and $2^{\rm nd}$ lower panels of Fig.~\ref{fig:debris}), and essentially all the bound mass is placed on highly-elliptical orbits with apocenters several thousand $r_{\rm g}$ in size ($3^{\rm rd}$ and $4^{\rm th}$ lower panels of Fig.~\ref{fig:debris}\footnote{ Data in the lower panels are taken from a simulation in which a $3M_\odot$ middle-aged main-sequence star on a parabolic orbit with $r_{\rm p} \simeq 110 r_{\rm g}$ is disrupted by a $10^5 M_\odot$ SMBH (Ryu et~al. in preparation).} ), and dissipative events within this bound debris power the photon flare.

Importantly, the  gas heating that powers the flare in eTDEs has a very different character from the stream-stream interactions seen in ordinary TDEs \citep[e.g.][]{Shiokawa+2015}. In eTDEs, shocks first form when the spirals merge into a shell, and then stronger shocks take place in the radially-infalling matter surrounding the SMBH. 
However, as illustrated in the $4^{\rm th}$ lower panel of Fig.~\ref{fig:debris}, in ordinary TDEs, shocks occur at specific intersection points between bound material moving on elliptical orbits, whether at a ``nozzle shock" stretching along a line whose inner end is at $\simeq r_{\rm p}$ \citep{Shiokawa+2015,SteinbergStone2022} or at ``apocenter shocks" taking place at a distance $\sim 100\times$ that of the infall shocks in eTDEs.
The shocks seen in eTDEs are also different from the discrete and isolated stream intersections envisioned as taking place when $r_p \lesssim 15r_{\rm g}$ \citep{LB2020,Batra+2021}.

\subsection{Energetics}
\label{subsec:dmde}

In an ordinary TDE, the distribution $dM/dE$ of debris mass with orbital energy $E\equiv -(u_{\rm t} + 1)$ is roughly a square wave with edges at  $\Delta E = \pm \Xi \Delta \epsilon$, where $\Delta \epsilon = G M_{\rm BH}^{1/3}M_*^{2/3}/R_* $, \citep{Rees1988} and $\Xi \approx 1 - 2$ \citep{Ryu1+2020}.
The top panel of Fig.~\ref{fig:B_distribution} displays both how different the immediately post-disruption $dM/dE$
is from that of an ordinary TDE  and how much it is redistributed well after the gas leaves the star. It also reveals how much of the bound debris is rapidly lost to accretion.

In an ordinary TDE, $dM/dE$ is symmetric around $E=0$,  nearly flat from $E=-\Delta E$ to $E = +\Delta E$, and drops sharply for $|E| > \Delta E$ \citep{Lodato2009,GR2013,Goicovic2019,Ryu2+2020}.
In an eTDE, by $t\sim$230~s after pericenter passage, although the energy distribution (in our fiducial simulation) is roughly symmetric, it is centered at $\simeq + (1 - 2)\Delta \epsilon$, and its half-width $\Delta E \sim 10~\Delta \epsilon$.
The debris energy distribution found by \citet{GaftonRosswog2019} at a similar time was qualitatively similar, but quantitatively different: narrower by a factor of a few and symmetric around $E\simeq 0$.   This contrast may result from our use of a main-sequence internal density profile rather than their $\gamma=5/3$ polytrope.

However, this distribution soon changes drastically, a change not seen in previous work because their calculations stopped before it begins.
Within $\sim 10$~minutes, most of the bound material plunges into the SMBH.   By $\sim 3$~hr, the radial pressure gradient within the spirals broadens the distribution of the remaining matter by a further factor $\sim 2-3$, while also making it highly asymmetric and decidedly not flat-topped (see Fig.~\ref{fig:B_distribution}, upper panel).

After the redistribution of energy, some of the bound material that had moved outward falls back toward the SMBH. The converging streams shock against each other, transforming orbital energy into heat. There it forms a compact ($\lesssim 100~r_{\rm g}$), hot (a few $10^{6}$ K), roughly spherical structure which is illustrated in the  inset in the $4^{\rm th}$ panel of Fig.~\ref{fig:debris}. The most tightly-bound matter enters this structure first; the sharp low-energy cut-off in $dM/dE$ at $t=40000$~s signals that the gas whose orbital energy had been $\simeq -(20 \, - \,30) \Delta\epsilon$ has moved to much more negative orbital energy due to dissipation in shocks.
Unlike a classic Keplerian accretion disk that is supported by angular momentum,  this accretion flow is geometrically thick and primarily radiation pressure-supported: the mean specific angular momentum is only about half what would be required for a circular orbit in this range of radii. 

Another consequence of the broad and asymmetric debris energy distribution is that the rate at which bound matter falls back toward the SMBH has a different time-dependence from that of ordinary TDEs.  Because $dM/dE$ rises with increasing $E$ steadily, but unevenly, across the entire range of bound energies, the post-peak decay of the mass fall-back rate declines
more slowly (see bottom panel of Fig.~\ref{fig:B_distribution}) than in the case of ordinary TDEs---crudely $\propto t^{-5/4}$ rather than $\propto t^{-5/3}$. However, as we discuss in Sec.~\ref{sec:luminosity}, as for ordinary TDEs (but for different reasons), the mass fallback rate does not translate directly into a lightcurve.

The unbound ejecta are contained in an axisymmetric ring of mass $\simeq 0.4M_\odot$ that moves continuously outward once it forms. It simultaneously expands vertically as radiation forces compete with gravity.
The distribution of outgoing speed at infinity can be estimated from the $dM/dE$ distribution (Fig.~\ref{fig:B_distribution}). For our fiducial case, the bulk of the unbound ejecta has specific orbital energy $\simeq (3-4)\Delta\epsilon$,
corresponding to a speed $\simeq 9000$~km/s at infinity. The total kinetic energy available
for deposition in the surrounding gas is $\simeq 10^{51}$~erg. However, a bit less than { $ 1\%$} of the ejecta mass has a speed $\gtrsim 21000$~km/s at infinity, a factor of $3-4$ 
faster than the same mass ejecta mass fraction for ordinary TDEs \citep{Ryu1+2020}. This fast expanding debris that carries $\sim 10^{50}$~erg can produce a strong flare, as discussed in Sec.~\ref{sec:radioflare} below.

\section{Observational implications}\label{sec:discussion}

\subsection{Luminosity and Spectrum}
\label{sec:luminosity}

The inner hot accretion flow is the main source of the radiation. 
We estimate the bolometric luminosity by integrating the local emissivity of cells within the photosphere whose cooling time is shorter than the evolution time.   So that the surface brightness may be used to define a characteristic spectral temperature, we use the thermalization photosphere, defined as the location at $\sqrt{\tau_{\rm T}\tau_{\rm ff}}\simeq 1$, where $\tau_{\rm T}$ is the Thomson optical depth and $\tau_{\rm ff}$ the absorption optical depth, both of which are integrated over polar angle. See Appendix ~\ref{appendix:luminosity} for details.

\begin{figure}
	\centering
 	\includegraphics[width=4cm,angle=0]{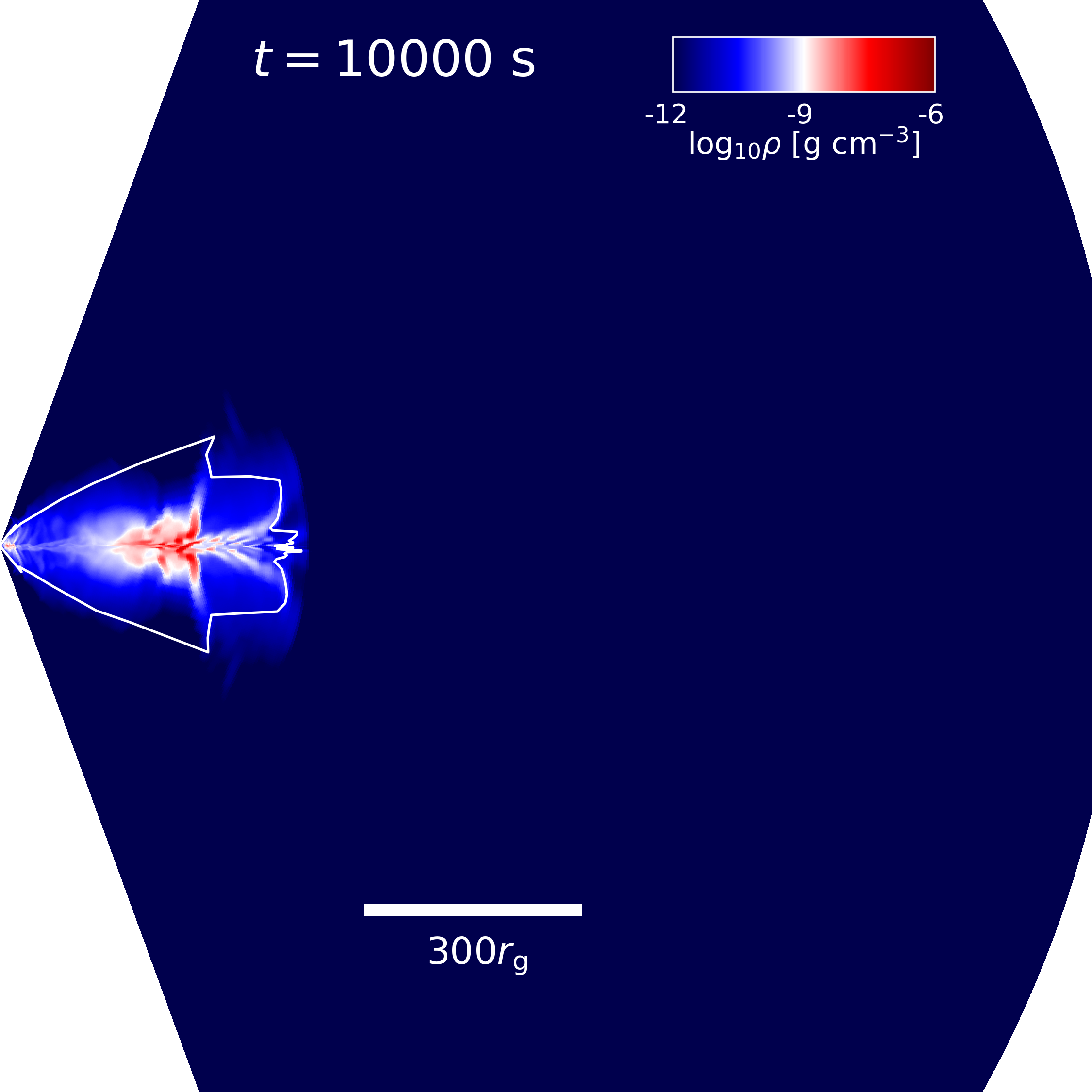}
	\includegraphics[width=4cm,angle=0]{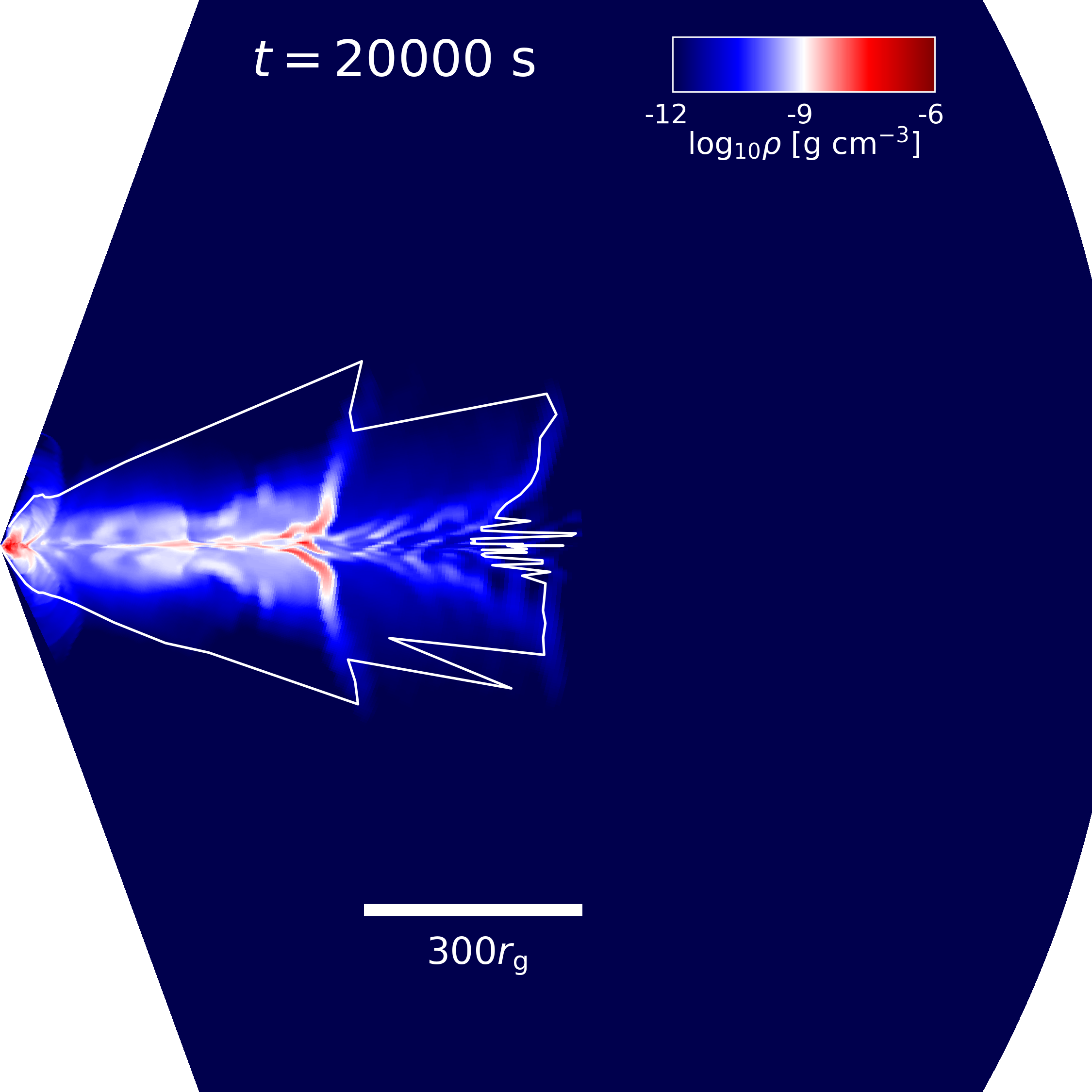}
	\includegraphics[width=4cm,angle=0]{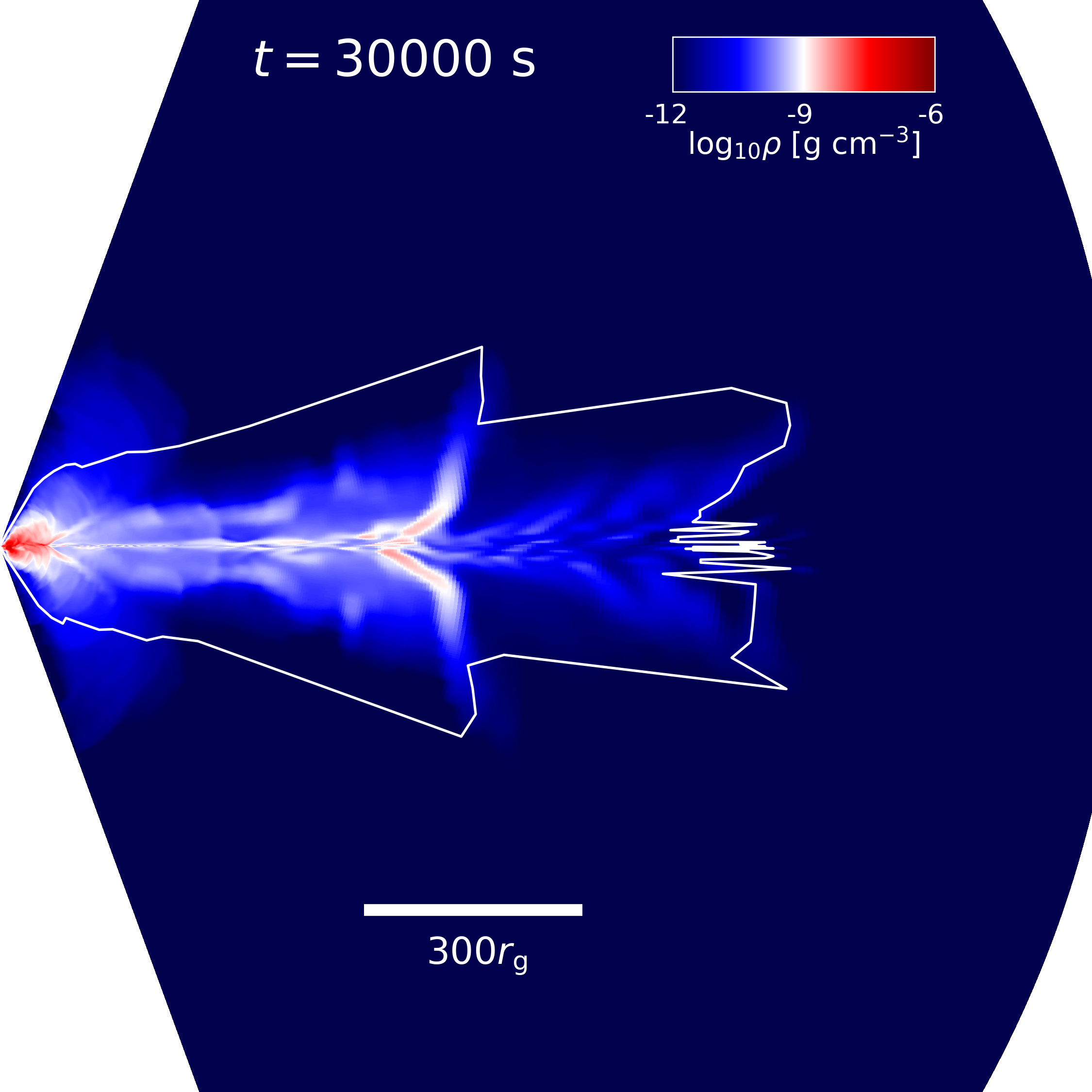}
	\includegraphics[width=4cm,angle=0]{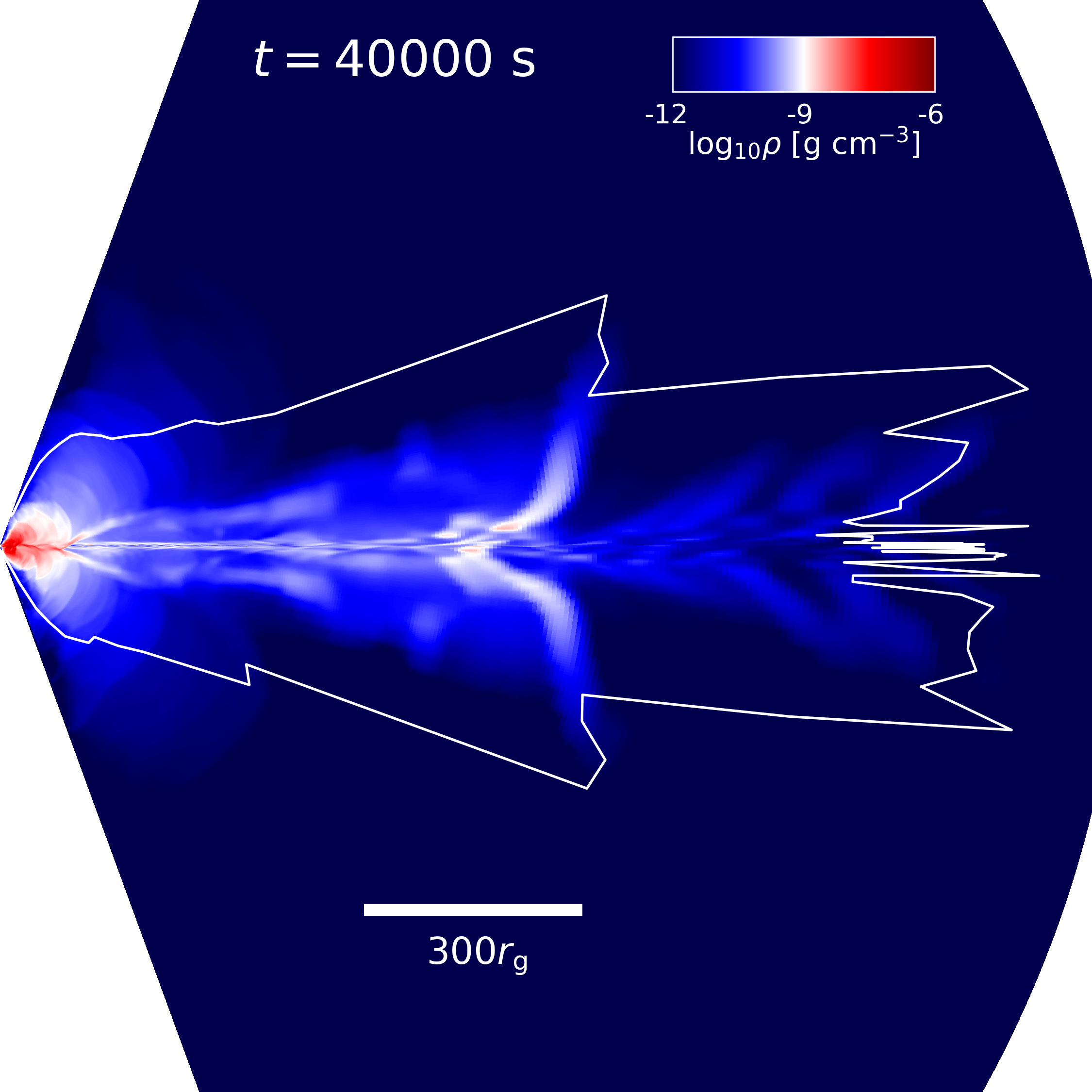} \\
	\includegraphics[width=4cm,angle=0]{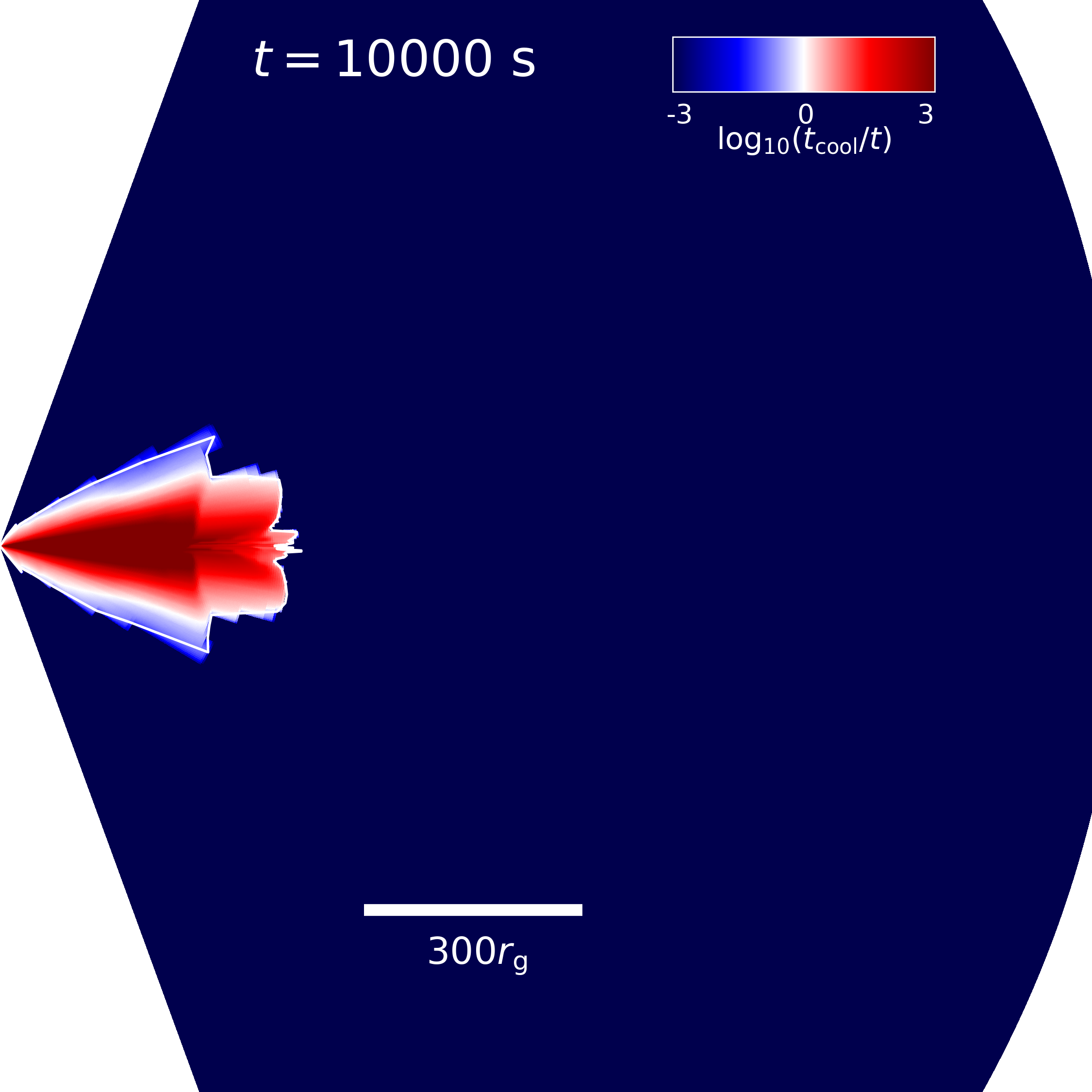}
	\includegraphics[width=4cm,angle=0]{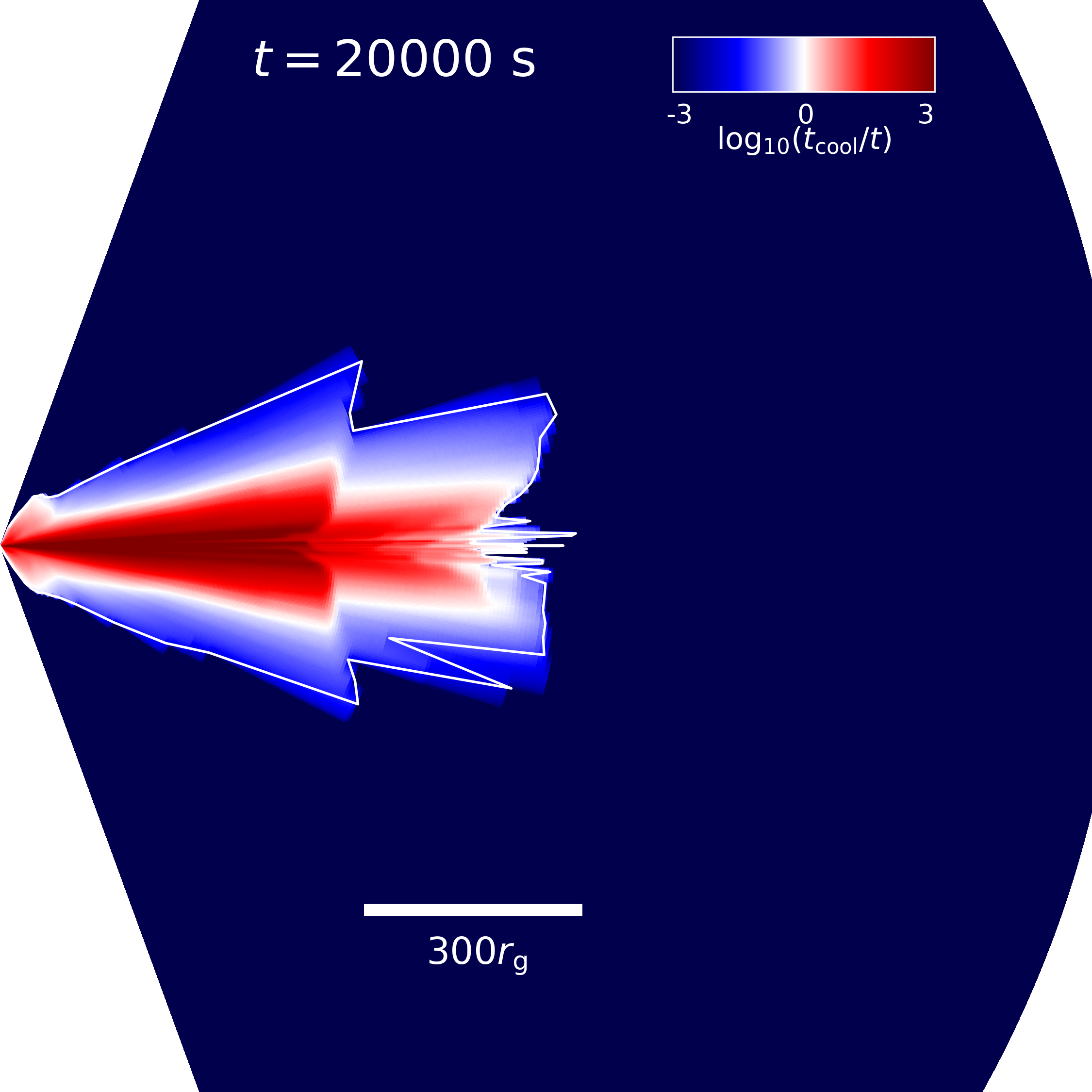}
	\includegraphics[width=4cm,angle=0]{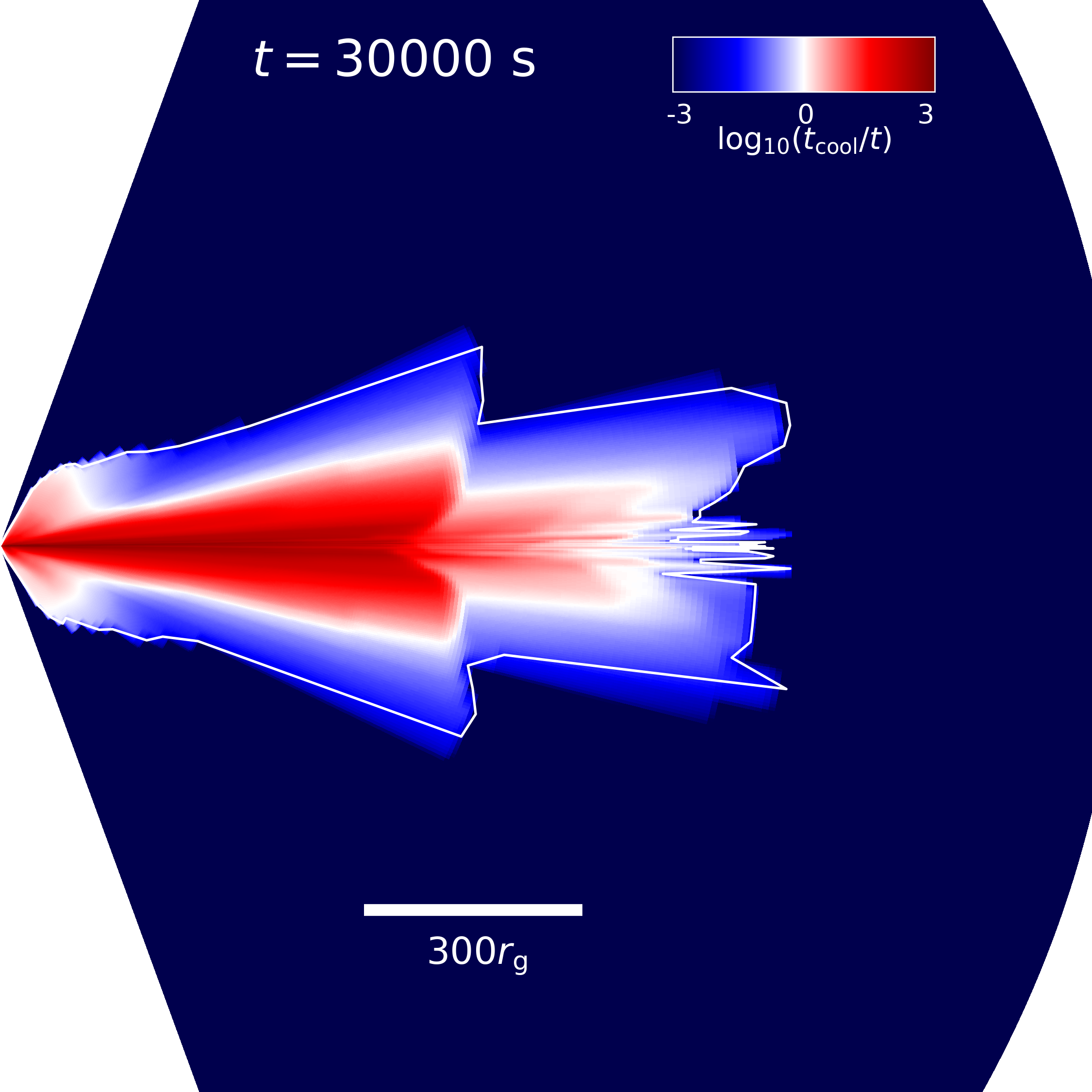}
	\includegraphics[width=4cm,angle=0]{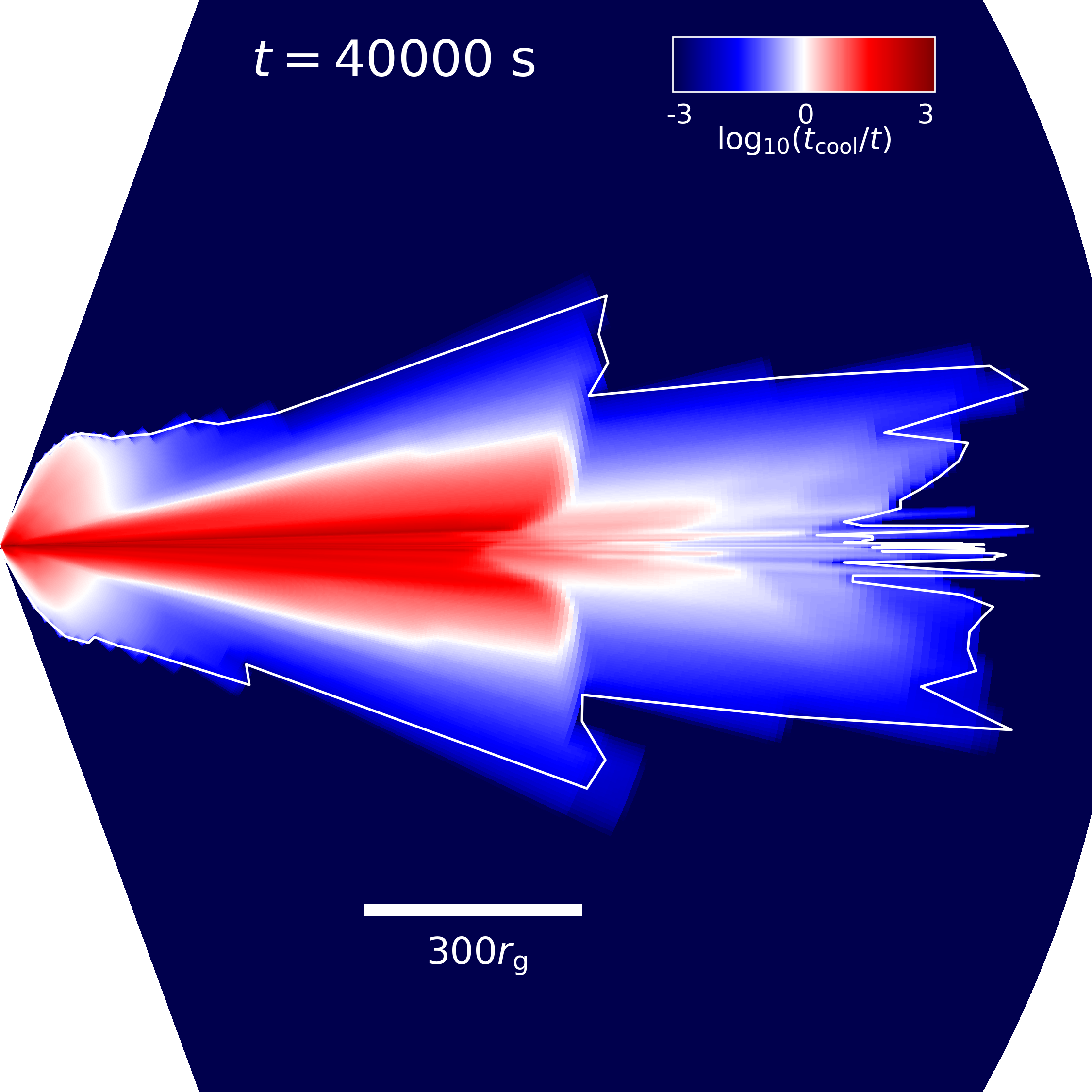}
\caption{The distribution of the density (\textit{top}) and the ratio of the cooling time to the evolution time (\textit{bottom})  in the $r-z$ plane at $\phi=0$ at $t\simeq10000$, 20000, 30000, 40000 s. The location of the thermalization photosphere as seen by a distant observer is plotted using a white line. We define the thermalization optical depth as $\sqrt{\tau_{\rm t} \tau_{\rm ff}}$, for $\tau_{\rm T}$ ($\tau_{\rm ff}$) the Thomson (absorption) optical depths integrated radially inwards from the outer boundary. The prominent break in structure roughly halfway out in the debris corresponds to the density peak of the expanding ring.}\label{fig:photosphere}
\end{figure}

The luminosity rises very rapidly---on a time scale of a few hours for the parameters of our simulations, rather than the $\sim 1$~month of ordinary TDEs,---and persists at roughly the Eddington luminosity ($\sim 2 - 3 \times 10^{44}$~erg~s$^{-1}$ for $M_{\rm BH} = 10^6 M_\odot$) until at least $\sim 1/2$~day, the time at the end of our simulations.  The speed of the lightcurve's rise can be seen in Figure~\ref{fig:photosphere}, which illustrates how rapidly the volume of the cooling region within the observed photosphere grows at the beginning of an event\footnote{Note that for this purpose, we define the photosphere by integrating the effective opacity along radial paths of differing polar angle.}.
At later times, the luminosity should persist at this level until the mass fallback rate becomes too small to support such a luminosity.  At that point, it should decline with the shallow power-law of the fallback rate. 
Our cooling time-based lightcurve estimate should capture the majority of the luminosity (the portion coming from the innermost region) reasonably well.  The luminosity from the outer regions, whose cooling time is the longest, is more uncertain. Our method tends to overestimate it, but, because it is already a minority contributor, this means the actual luminosity may be less than estimated, but not by much.  Future time-dependent transfer studies will clarify this situation.

To characterize the spectrum, we calculate the effective temperature for each surface element of the photosphere using the local area and the local luminosity. We find that the effective temperature distribution is well-described by a single peak at $\sim 10^6$~K. Thus, the power is primarily in soft X-rays. For events driven by SMBHs of different masses, the luminosity peak should scale like the Eddington luminosity, $\propto M_{\rm BH}$, while the temperature is $\propto M_{\rm BH}^{-1/4}$ and the duration is $\propto M_* M_{\rm BH}^{-1}$.  This last scaling follows from the fact that the emitted energy is $\propto M_{\star}$ (see next subsection), but nearly independent of $M_{\rm BH}$,
while the Eddington luminosity is $\propto M_{\rm BH}$.

Although the peak luminosity is comparable to Eddington, there is relatively little matter far from the flow; consequently, reprocessing should be minor.  To demonstrate this, in Figure~\ref{fig:photosphere} we show the shape of the thermalization photosphere  as seen by distant observers at several times; we define the photosphere by integrating the opacity inward from the outer boundary along radial paths.  It is roughly axisymmetric everywhere; t is nearly flat close to the BH, e.g.,  at a distance $\simeq 100 r_{\rm g}$ from the equatorial plane for all radii $\gtrsim 100 r_{\rm g}$ at $t=40000$ s. As a result, the character of the continuum spectrum is determined fairly close to the site of initial radiation, the primary direction of photon diffusion is perpendicular to the orbital plane, and very little light emerges in directions close to the orbital plane.

The peak luminosity, spectrum, and lightcurve of eTDEs are therefore very different from those of ordinary TDEs, which are mostly observed in the optical with a peak luminosity lower by an order of magnitude, a rise time of order a month,
and a post-peak luminosity falling as a steep power-law in time.  
Although eTDEs' expected spectral shape in the X-ray band resembles that of ordinary TDEs (i.e., thermal with $T \sim 10^6$~K), the X-ray lightcurves of ordinary TDEs are qualitatively similar to those seen in the optical band \citep{Auchettl+2017,vanVelzen+2019,Jonker+2020}, rising comparatively gradually and then declining, rather than the extremely rapid rise to a plateau expected in eTDEs.  Even more tellingly, the observed X-ray luminosities of ordinary TDEs ($\sim 10^{42}$ -- $10^{43}$~erg~s$^{-1}$: \citealt{Auchettl+2017}) are much lower than the Eddington-limited luminosities we predict for eTDEs\footnote{There are three cases presented in \cite{Auchettl+2017} with higher X-ray luminosities, but their nature is uncertain because each has only a single observation.}.

\subsection{Total radiated energy}

Within the duration of our simulation, the luminosity estimated using the local cooling time sums to a total energy $\sim 2 \times 10^{48} (M_{\star}/M_\odot)$~erg. To estimate the radiated energy at later times, we first note that
the total energy available is $\sim \eta_{\rm diss} \Delta M c^2$, where $\eta_{\rm diss}$ is the energy per unit mass acquired by the radiating debris from dissipative processes and $\Delta M$ is the amount of remaining bound mass in the hot compact settling flow.  From shocks in this flow taking place at $\sim 50 r_{\rm g}$, $\eta_{\rm diss} \simeq 0.02$.
The total bound mass is $\Delta M/M_\odot \simeq 0.15 M_\odot$, suggesting
the total amount of energy radiated during the event might rise to $\sim 5\times 10^{51}$~erg or more.  Radiated at the Eddington luminosity, the luminosity we have estimated at the end of our simulation, such a flare would last $\sim 1 (M_{\star}/1 M_\odot)(M_{\rm BH}/10^6 M_\odot)^{-1}$~yr.

\subsection{Radio Flare}
\label{sec:radioflare}

The interaction of the  expanding  ejecta with the surrounding gas should produce a  radio flare \citep{Krolik+2016,Yalinewich+2019,Matsumoto+2021}.  As discussed in the TDE context by \citet{Krolik+2016}, electron acceleration at the shock driven by the ejecta leads to synchrotron emission whose peak flux depends on the ejecta velocity as: 
$F_\nu\propto f_A^{2/7} f_V^{5/7} v^{\frac{58-19k}{14}}$,
where we assume the energy distribution of the emitting electrons is $\propto E_e^{-3}$, as is typical for Newtonian shocks.  We further assume that the external density declines with distance  $ \propto r^{-k}$ . The factors $f_A$ and $f_V$ describe the area and volume of the emitting region as compared with those of a spherical outflow \citep{Barniol2013}. 
Because eTDEs have both unbound material with larger velocity and larger covering factors $f_A$ and $f_V$, their characteristic radio signal should be larger by  an order of magnitude compared to that of an ordinary TDE with the same external density.
This stronger radio flare could help identifying eTDEs in addition to their strong earlier X-ray signature.

Initially the  peak radio flux varies with time $\propto t^{19(2-k)/14}$ \citep{Krolik+2016}.  For a gradually declining density profile, like in the Milky Way (where $k \approx 1$), the radio luminosity increases with time. For steeper density profiles, like those observed in most TDE hosts, which generally have $k \approx 2$ \citep{Matsumoto+2021}, the flux is roughly constant.  In either case the peak frequency decreases with time until eventually it  drops significantly below  1~GHz and the source becomes undetectable.  Overall, as a significant fraction of the unbound material moves at $\gtrsim 20\,000$ km/s (see \S \ref{subsec:dmde}), which is much faster than a regular TDE, the radio signal should be brighter by about a factor of $\sim 10$  and longer by a factor of $\sim 3$ than the
radio emission of ordinary (unjetted) TDEs.

\begin{figure}
	\centering
	\includegraphics[height=6.5cm,angle=0]{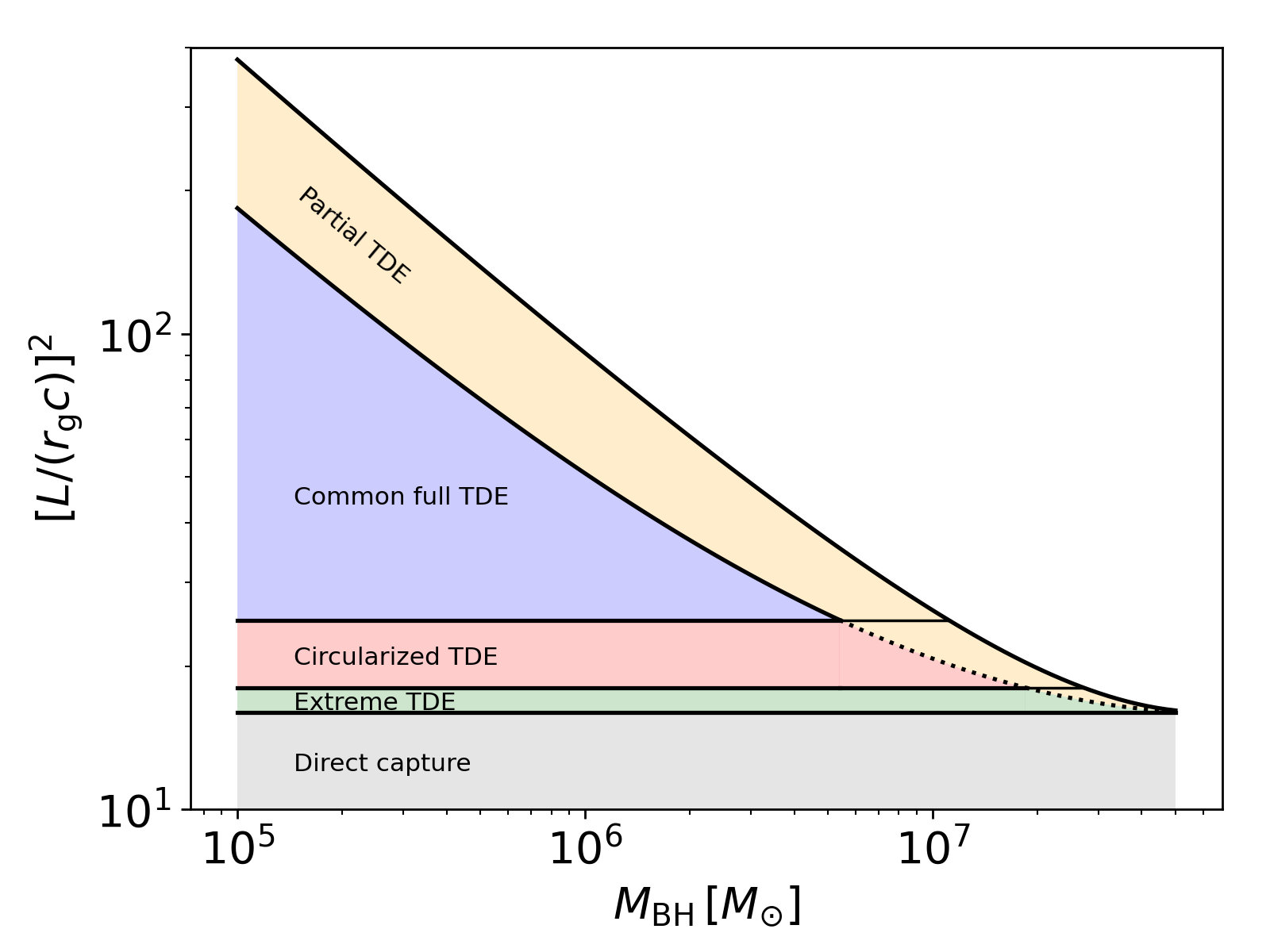}
	\includegraphics[height=6.5cm,angle=0]{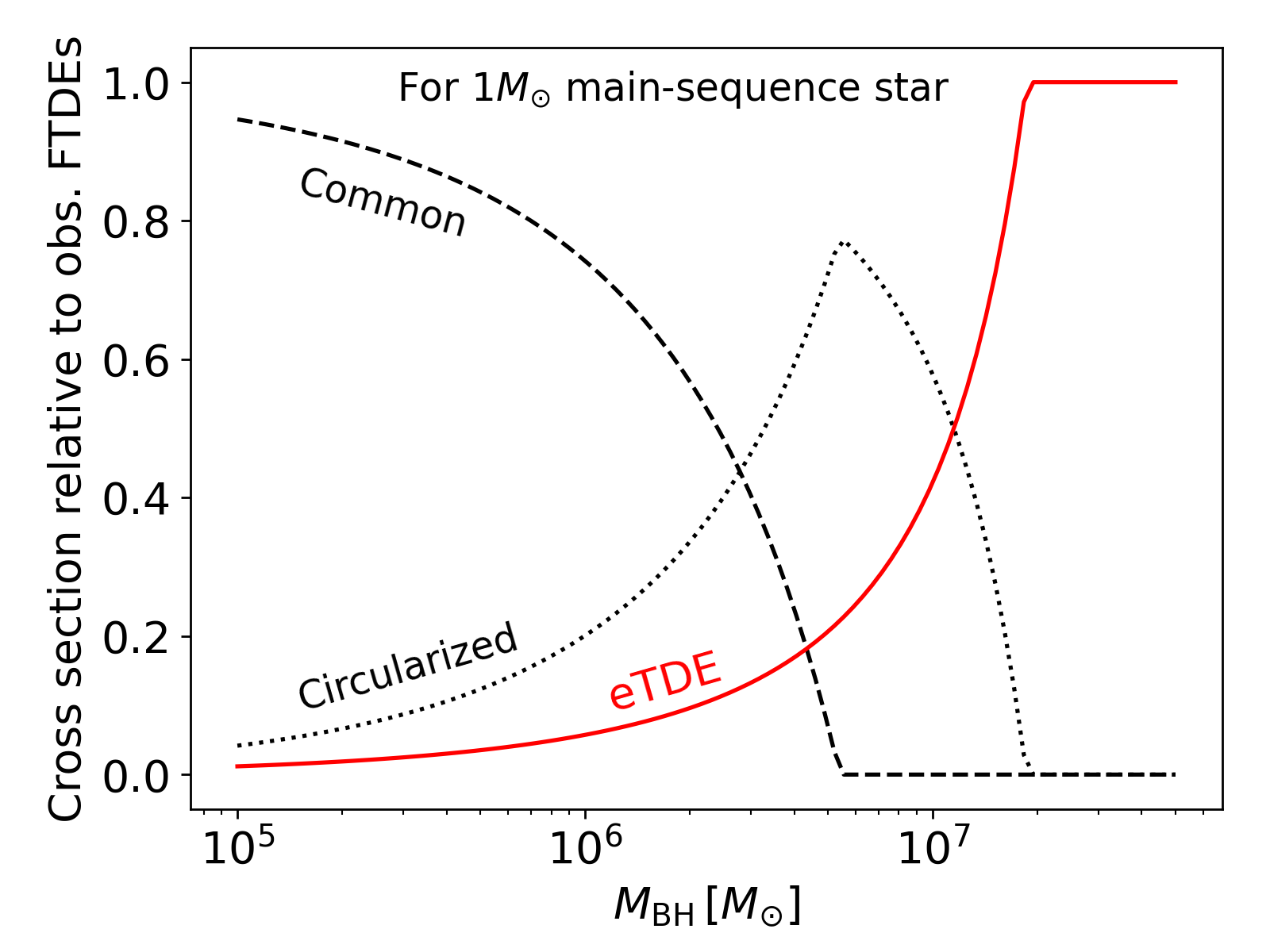} 
	\caption{(\textit{top}) Regions in parameter space
	for five kinds of disruption events: partial TDE (orange), common full TDE (blue), circularized TDE (red),
    and direct capture (gray, $r_{\rm p}<4_{\rm g}$). 
	We define partial TDEs as events where the star loses more than 10\% of its mass at the first pericenter passage. This plot is an extended version of Fig.~3 in \citet{Krolik+2020}. (\textit{bottom}) The cross section as a function of $M_{\rm BH}$ for each class of observable full disruption events, i.e., common (dashed black), circularized (dotted black), extreme (solid red), in units of the total disruption cross section for a $1\Msol$ main-sequence star.   }\label{fig:crosssection}
\end{figure}

\subsection{Rate}
 For a fixed stellar distribution function that varies little across the loss-cone, the rate of events having a pericenter less than $r_p$ (measured in units of $r_g$) is $\propto L^2(r_p)$, which is $2r_p^2/(r_p -2)$ in Schwarzschild spacetime.
To illustrate how the relative rates of different varieties of TDEs depend on $M_{\rm BH}$, we show in the \textit{top} panel of Fig.~\ref{fig:crosssection} $L^{2}$ for partial TDEs, common full TDEs, circularized TDEs, eTDEs and direct capture events\footnote{Following \citet{Krolik+2020}, we define circularized TDEs to be events with strong enough relativistic apsidal precession for debris to circularize rapidly.} and in the \textit{bottom} panel the cross sections of the different varieties of observable full disruption events relative to their total. 
When the black hole mass is relatively small ($10^6 M_\odot$), the rate of eTDEs is only $\simeq 6\%$ of all observable events (i.e., excluding direct captures). This fraction is, nonetheless, only about a factor of 3 smaller than that of circularized events for this black hole mass.
However, as $M_{\rm BH}$ increases, these extreme events become a much larger fraction of all those displaying observable signals: for $M_{\rm BH}\gtrsim 10^{7}M_{\odot}$, they are $\gtrsim 40\%$ of all observable TDEs, becoming the majority for $M_{\rm BH} \gtrsim (2-4) \times 10^7 M_\odot$ (depending on the stellar mass).
 If we apply these relative rates to the observationally calibrated TDE rate estimated by \citet{StoneMetzger2016}, the rate of eTDEs as a function of $M_{\rm BH}$ peaks at $M_{\rm BH}\simeq 2\times 10^{7}\Msol$, where the rate is $\approx 6\times10^{-5}$ yr$^{-1}$ per galaxy.

\section{Summary and conclusion}\label{sec:summary}

In this paper, we examined the long-term evolution of debris produced in extremely relativistic tidal disruption events of a realistic main-sequence $1\Msol$ star by a $10^{6}\Msol$ SMBH using fully relativistic hydrodynamics simulations with realistic initial conditions. We considered several different pericenter distances ranging from $r_{\rm p}\simeq 4.03- 7 r_{\rm g}$. Strikingly, extremely strong apsidal precession, which occurs only for $r_{\rm p}\lesssim 6 r_{\rm g}$, leads to a debris evolution qualitatively different from that for ordinary disruption events. The debris undergoes four different phases: it is elongated to form a crescent that  stretches to form a spiral wrapping around the SMBH. The spiral expands outward, and then its several windings merge into a ring that continues to advance outwards.  The bound material of the ring eventually falls back toward the SMBH and forms a roughly spherical accretion flow near the SMBH. The resulting hot ($\sim (1-2)\times 10^6$~K) accretion flow is the main source of radiation. Our detailed analysis indicates that the luminosity rises on a time scale ($\sim 3$~hr for $M_{\rm BH}\sim 10^6 M_\odot$) much shorter than the flare duration ( $\sim 1 \, (M_{\rm BH}/10^6 M_\odot)^{-1}$~yr). For most of its duration, the flare should maintain approximately the Eddington luminosity of the SMBH, but then decline as a shallow power-law when the continued infall cannot sustain that luminosity.

These events should be detectable by eROSITA as X-ray events that are accompanied by weak or no optical signal. The interaction of the high velocity ($\gtrsim 20000$~km/s) escaping unbound material with the surrounding matter should lead to a powerful radio flare \citep{Krolik+2016,Yalinewich+2019,Matsumoto+2021} that follows these events by a few weeks.

Thus, even though they are genuine tidal disruptions, their lightcurves and spectra are very different from classical expectations; consequently, matching the classical expectations should not be an absolute prerequisite for classification as a TDE.
Although these events are probably rare for lower mass SMBHs (i.e., $M_{\rm BH}\lesssim 10^{6}M_{\odot}$), they should be the dominant tidal disruption events yielding flares for $M_{\rm BH}\gtrsim 3\times 10^{7}M_{\odot}$.
 Moreover, any such event in a SMBH of this mass would be exceptionally luminous because $L_{\rm E}\simeq 4.5 \times 10^{45}(M_{\rm BH}/3\times 10^{7}M_{\odot})$~erg~s$^{-1}$.

\section*{Acknowledgments}
This research project was conducted using computational resources (and/or scientific computating services) at the Max-Planck Computing \& Data Facility and at the Maryland Advanced Research
 Computing Center (MARCC). The simulations were performed on the national supercomputer Hawk at the High Performance Computing Center Stuttgart (HLRS) under the grant number TDEglobalsimulation/44232. J.H.K. was partially supported by NSF grant AST-2009260. T.P. was partially supported by ERC advanced grants TReX and MultiJets. 

\software{
matplotlib \citep{Hunter:2007}; \mesa ~\citep{Paxton+2011}; 
\harm ~ \citep{Noble+2009}.
}
\bibliography{biblio}

\begin{thebibliography}{}
	\expandafter\ifx\csname natexlab\endcsname\relax\def\natexlab#1{#1}\fi
	\providecommand{\url}[1]{\href{#1}{#1}}
	\providecommand{\dodoi}[1]{doi:~\href{http://doi.org/#1}{\nolinkurl{#1}}}
	\providecommand{\doeprint}[1]{\href{http://ascl.net/#1}{\nolinkurl{http://ascl.net/#1}}}
	\providecommand{\doarXiv}[1]{\href{https://arxiv.org/abs/#1}{\nolinkurl{https://arxiv.org/abs/#1}}}
	
	\bibitem[{{Auchettl} {et~al.}(2017){Auchettl}, {Guillochon}, \&
		{Ramirez-Ruiz}}]{Auchettl+2017}
	{Auchettl}, K., {Guillochon}, J., \& {Ramirez-Ruiz}, E. 2017, \apj, 838, 149,
	\dodoi{10.3847/1538-4357/aa633b}
	
	\bibitem[{{Barniol Duran} {et~al.}(2013){Barniol Duran}, {Nakar}, \&
		{Piran}}]{Barniol2013}
	{Barniol Duran}, R., {Nakar}, E., \& {Piran}, T. 2013, \apj, 772, 78,
	\dodoi{10.1088/0004-637X/772/1/78}
	
	\bibitem[{{Batra} {et~al.}(2021){Batra}, {Lu}, {Bonnerot}, \&
		{Phinney}}]{Batra+2021}
	{Batra}, G., {Lu}, W., {Bonnerot}, C., \& {Phinney}, E.~S. 2021, arXiv
	e-prints, arXiv:2112.03918.
	\newblock \doarXiv{2112.03918}
	
	\bibitem[{{Carter} \& {Luminet}(1983)}]{CarterLuminet}
	{Carter}, B., \& {Luminet}, J.~P. 1983, \aap, 121, 97
	
	\bibitem[{Cheng \& Bogdanovi\ifmmode~\acute{c}\else
		\'{c}\fi{}(2014)}]{Cheng2014}
	Cheng, R.~M., \& Bogdanovi\ifmmode~\acute{c}\else \'{c}\fi{}, T. 2014, Phys.
	Rev. D, 90, 064020, \dodoi{10.1103/PhysRevD.90.064020}
	
	\bibitem[{{Darbha} {et~al.}(2019){Darbha}, {Coughlin}, {Kasen}, \&
		{Nixon}}]{Darbha+2019}
	{Darbha}, S., {Coughlin}, E.~R., {Kasen}, D., \& {Nixon}, C. 2019, \mnras, 488,
	5267, \dodoi{10.1093/mnras/stz1923}
	
	\bibitem[{{Evans} {et~al.}(2015){Evans}, {Laguna}, \& {Eracleous}}]{Evans+2015}
	{Evans}, C., {Laguna}, P., \& {Eracleous}, M. 2015, \apjl, 805, L19,
	\dodoi{10.1088/2041-8205/805/2/L19}
	
	\bibitem[{{Gafton} \& {Rosswog}(2019)}]{GaftonRosswog2019}
	{Gafton}, E., \& {Rosswog}, S. 2019, \mnras, 487, 4790,
	\dodoi{10.1093/mnras/stz1530}
	
	\bibitem[{{Gezari}(2021)}]{Gezari2021}
	{Gezari}, S. 2021, \araa, 59, \dodoi{10.1146/annurev-astro-111720-030029}
	
	\bibitem[{{Goicovic} {et~al.}(2019){Goicovic}, {Springel}, {Ohlmann}, \&
		{Pakmor}}]{Goicovic2019}
	{Goicovic}, F.~G., {Springel}, V., {Ohlmann}, S.~T., \& {Pakmor}, R. 2019,
	\mnras, 487, 981, \dodoi{10.1093/mnras/stz1368}
	
	\bibitem[{{Guillochon} \& {Ramirez-Ruiz}(2013)}]{GR2013}
	{Guillochon}, J., \& {Ramirez-Ruiz}, E. 2013, \apj, 767, 25,
	\dodoi{10.1088/0004-637X/767/1/25}
	
	\bibitem[{{Halpern} {et~al.}(2004){Halpern}, {Gezari}, \&
		{Komossa}}]{Halpern+2004}
	{Halpern}, J.~P., {Gezari}, S., \& {Komossa}, S. 2004, \apj, 604, 572,
	\dodoi{10.1086/381937}
	
	\bibitem[{{Hung} {et~al.}(2017){Hung}, {Gezari}, {Blagorodnova}, {Roth},
		{Cenko}, {Kulkarni}, {Horesh}, {Arcavi}, {McCully}, {Yan}, {Lunnan},
		{Fremling}, {Cao}, {Nugent}, \& {Wozniak}}]{Hung+2017}
	{Hung}, T., {Gezari}, S., {Blagorodnova}, N., {et~al.} 2017, \apj, 842, 29,
	\dodoi{10.3847/1538-4357/aa7337}
	
	\bibitem[{Hunter(2007)}]{Hunter:2007}
	Hunter, J.~D. 2007, Computing In Science \& Engineering, 9, 90,
	\dodoi{10.1109/MCSE.2007.55}
	
	\bibitem[{{Iglesias} \& {Rogers}(1996)}]{OPAL}
	{Iglesias}, C.~A., \& {Rogers}, F.~J. 1996, \apj, 464, 943,
	\dodoi{10.1086/177381}
	
	\bibitem[{{Jonker} {et~al.}(2020){Jonker}, {Stone}, {Generozov}, {van Velzen},
		\& {Metzger}}]{Jonker+2020}
	{Jonker}, P.~G., {Stone}, N.~C., {Generozov}, A., {van Velzen}, S., \&
	{Metzger}, B. 2020, \apj, 889, 166, \dodoi{10.3847/1538-4357/ab659c}
	
	\bibitem[{{Kobayashi} {et~al.}(2004){Kobayashi}, {Laguna}, {Phinney}, \&
		{M{\'e}sz{\'a}ros}}]{Kobayashi+2004}
	{Kobayashi}, S., {Laguna}, P., {Phinney}, E.~S., \& {M{\'e}sz{\'a}ros}, P.
	2004, \apj, 615, 855, \dodoi{10.1086/424684}
	
	\bibitem[{{Komossa} \& {Bade}(1999)}]{KomossaBade1999}
	{Komossa}, S., \& {Bade}, N. 1999, \aap, 343, 775.
	\newblock \doarXiv{astro-ph/9901141}
	
	\bibitem[{{Kormendy} \& {Ho}(2013)}]{KormendyHoe2013}
	{Kormendy}, J., \& {Ho}, L.~C. 2013, \araa, 51, 511,
	\dodoi{10.1146/annurev-astro-082708-101811}
	
	\bibitem[{{Krolik} {et~al.}(2020){Krolik}, {Piran}, \& {Ryu}}]{Krolik+2020}
	{Krolik}, J., {Piran}, T., \& {Ryu}, T. 2020, \apj, 904, 68,
	\dodoi{10.3847/1538-4357/abc0f6}
	
	\bibitem[{{Krolik} {et~al.}(2016){Krolik}, {Piran}, {Svirski}, \&
		{Cheng}}]{Krolik+2016}
	{Krolik}, J., {Piran}, T., {Svirski}, G., \& {Cheng}, R.~M. 2016, \apj, 827,
	127, \dodoi{10.3847/0004-637X/827/2/127}
	
	\bibitem[{{Krolik}(2010)}]{Krolik2010}
	{Krolik}, J.~H. 2010, \apj, 709, 774, \dodoi{10.1088/0004-637X/709/2/774}
	
	\bibitem[{{Lacy} {et~al.}(1982){Lacy}, {Townes}, \& {Hollenbach}}]{Lacy+}
	{Lacy}, J.~H., {Townes}, C.~H., \& {Hollenbach}, D.~J. 1982, \apj, 262, 120,
	\dodoi{10.1086/160402}
	
	\bibitem[{{Laguna} {et~al.}(1993){Laguna}, {Miller}, {Zurek}, \&
		{Davies}}]{Laguna+1993}
	{Laguna}, P., {Miller}, W.~A., {Zurek}, W.~H., \& {Davies}, M.~B. 1993, \apjl,
	410, L83, \dodoi{10.1086/186885}
	
	\bibitem[{{Lodato} {et~al.}(2009){Lodato}, {King}, \& {Pringle}}]{Lodato2009}
	{Lodato}, G., {King}, A.~R., \& {Pringle}, J.~E. 2009, \mnras, 392, 332,
	\dodoi{10.1111/j.1365-2966.2008.14049.x}
	
	\bibitem[{{Lu} \& {Bonnerot}(2020)}]{LB2020}
	{Lu}, W., \& {Bonnerot}, C. 2020, \mnras, 492, 686,
	\dodoi{10.1093/mnras/stz3405}
	
	\bibitem[{{Matsumoto} \& {Piran}(2021)}]{Matsumoto+2021}
	{Matsumoto}, T., \& {Piran}, T. 2021, \mnras, 507, 4196,
	\dodoi{10.1093/mnras/stab2418}
	
	\bibitem[{{Noble} {et~al.}(2009){Noble}, {Krolik}, \& {Hawley}}]{Noble+2009}
	{Noble}, S.~C., {Krolik}, J.~H., \& {Hawley}, J.~F. 2009, \apj, 692, 411,
	\dodoi{10.1088/0004-637X/692/1/411}
	
	\bibitem[{{Paxton} {et~al.}(2011){Paxton}, {Bildsten}, {Dotter}, {Herwig},
		{Lesaffre}, \& {Timmes}}]{Paxton+2011}
	{Paxton}, B., {Bildsten}, L., {Dotter}, A., {et~al.} 2011, \apjs, 192, 3,
	\dodoi{10.1088/0067-0049/192/1/3}
	
	\bibitem[{{Phinney}(1989)}]{Phinney1989}
	{Phinney}, E.~S. 1989, in The Center of the Galaxy, ed. M.~{Morris}, Vol. 136,
	543
	
	\bibitem[{{Piran} {et~al.}(2015){Piran}, {Svirski}, {Krolik}, {Cheng}, \&
		{Shiokawa}}]{Piran2015}
	{Piran}, T., {Svirski}, G., {Krolik}, J., {Cheng}, R.~M., \& {Shiokawa}, H.
	2015, \apj, 806, 164, \dodoi{10.1088/0004-637X/806/2/164}
	
	\bibitem[{{Rees}(1988)}]{Rees1988}
	{Rees}, M.~J. 1988, \nat, 333, 523, \dodoi{10.1038/333523a0}
	
	\bibitem[{{Ryu} {et~al.}(2020{\natexlab{a}}){Ryu}, {Krolik}, {Piran}, \&
		{Noble}}]{Ryu1+2020}
	{Ryu}, T., {Krolik}, J., {Piran}, T., \& {Noble}, S.~C. 2020{\natexlab{a}},
	\apj, 904, 98, \dodoi{10.3847/1538-4357/abb3cf}
	
	\bibitem[{{Ryu} {et~al.}(2020{\natexlab{b}}){Ryu}, {Krolik}, {Piran}, \&
		{Noble}}]{Ryu2+2020}
	---. 2020{\natexlab{b}}, \apj, 904, 99, \dodoi{10.3847/1538-4357/abb3cd}
	
	\bibitem[{{Shiokawa} {et~al.}(2015){Shiokawa}, {Krolik}, {Cheng}, {Piran}, \&
		{Noble}}]{Shiokawa+2015}
	{Shiokawa}, H., {Krolik}, J.~H., {Cheng}, R.~M., {Piran}, T., \& {Noble}, S.~C.
	2015, \apj, 804, 85, \dodoi{10.1088/0004-637X/804/2/85}
	
	\bibitem[{{Steinberg} \& {Stone}(2022)}]{SteinbergStone2022}
	{Steinberg}, E., \& {Stone}, N.~C. 2022, arXiv e-prints, arXiv:2206.10641.
	\newblock \doarXiv{2206.10641}
	
	\bibitem[{{Stone} \& {Metzger}(2016)}]{StoneMetzger2016}
	{Stone}, N.~C., \& {Metzger}, B.~D. 2016, \mnras, 455, 859,
	\dodoi{10.1093/mnras/stv2281}
	
	\bibitem[{{Svirski} {et~al.}(2017){Svirski}, {Piran}, \&
		{Krolik}}]{Svirski2017}
	{Svirski}, G., {Piran}, T., \& {Krolik}, J. 2017, \mnras, 467, 1426,
	\dodoi{10.1093/mnras/stx117}
	
	\bibitem[{{Tejeda} {et~al.}(2017){Tejeda}, {Gafton}, {Rosswog}, \&
		{Miller}}]{Tejeda+2017}
	{Tejeda}, E., {Gafton}, E., {Rosswog}, S., \& {Miller}, J.~C. 2017, \mnras,
	469, 4483, \dodoi{10.1093/mnras/stx1089}
	
	\bibitem[{{van Velzen} {et~al.}(2019){van Velzen}, {Stone}, {Metzger},
		{Gezari}, {Brown}, \& {Fruchter}}]{vanVelzen+2019}
	{van Velzen}, S., {Stone}, N.~C., {Metzger}, B.~D., {et~al.} 2019, \apj, 878,
	82, \dodoi{10.3847/1538-4357/ab1844}
	
	\bibitem[{{van Velzen} {et~al.}(2021){van Velzen}, {Gezari}, {Hammerstein},
		{Roth}, {Frederick}, {Ward}, {Hung}, {Cenko}, {Stein}, {Perley}, {Taggart},
		{Foley}, {Sollerman}, {Blagorodnova}, {Andreoni}, {Bellm}, {Brinnel}, {De},
		{Dekany}, {Feeney}, {Fremling}, {Giomi}, {Golkhou}, {Graham}, {Ho},
		{Kasliwal}, {Kilpatrick}, {Kulkarni}, {Kupfer}, {Laher}, {Mahabal}, {Masci},
		{Miller}, {Nordin}, {Riddle}, {Rusholme}, {van Santen}, {Sharma}, {Shupe}, \&
		{Soumagnac}}]{vanVelzen+2021}
	{van Velzen}, S., {Gezari}, S., {Hammerstein}, E., {et~al.} 2021, \apj, 908, 4,
	\dodoi{10.3847/1538-4357/abc258}
	
	\bibitem[{{Yalinewich} {et~al.}(2019){Yalinewich}, {Steinberg}, {Piran}, \&
		{Krolik}}]{Yalinewich+2019}
	{Yalinewich}, A., {Steinberg}, E., {Piran}, T., \& {Krolik}, J.~H. 2019,
	\mnras, 487, 4083, \dodoi{10.1093/mnras/stz1567}
	
\end{thebibliography}


\section{Numerical grid}\label{appendix:numericalgrid}

In the simulations with a spherical grid, we adopt modified spherical coordinates in Schwarzschild spacetime: the modified spherical coordinate variables ($r'$, $\theta'$, $\phi'$) are related to ordinary spherical coordinates ($r$, $\theta$, $\phi$) by,
\begin{align}\label{eq:coord1}
	r &= e^{r'},\\
	\theta & = \theta_{0} (\tanh[b(\theta' - a)] + \tanh[b(\theta' + a)]) + 0.5\pi,\\
	\phi &= \phi'.
\end{align}
Here, $\theta_{0}= -(0.5\pi- \theta_{c})/[\tanh(b(-0.5 - a))+\tanh(b(-0.5 + a))]$.  The angle $\theta_{c}$ is the opening angle of the polar cut-out, and $a$ and $b$ are a set of tuning parameters that determine the vertical structure, which are given within $0.32\leq a\leq 0.35$ and $9.8\leq b \leq 10$. These modified coordinates allow us to place the grid cells where they are most needed in the simulation domain. The radial grid cells have constant $\Delta r / r$ and the vertical cells are more concentrated towards the mid-plane. To minimize the computational cost, we flexibly adjust the domain extent in $\theta$ and $r$. During the grid transition, we adjust the number of cells to ensure that there are more than 15-20 cells per scale height in $r$, $\theta$ and $\phi$.

 The boundary conditions are  outflow for the $r$ and $\theta$ boundaries and periodic for the $\phi$ boundary. The Courant number is 0.3.

\section{Luminosity estimate}\label{appendix:luminosity}

Because our simulations do not include time-dependent radiation transfer, we estimate the luminosity based on the local cooling time. 
Here, we define the local cooling time as $t_{\rm cool}=h_{\rho} \tau( 1+ u_{\rm gas}/u_{\rm rad})/ c $ where $h_{\rho}$ is the density scale height along the $\theta-$direction,  $\tau$ is the optical depth integrated along $\theta$-coordinate curves from the polar angle cut-out to the individual cells and $u_{\rm gas}$ ($u_{\rm rad}$) is the local gas thermal (radiation) energy contained within the cell. The opacity is found in terms of $\rho$ and $T$ using an OPAL opacity table for Solar metallicity \citep{OPAL}.

At early evolutionary stages ($t\lesssim $ a few hours, where $t$ is the time since pericenter passage), the gas is packed into dense spirals that then merge into an expanding ring.  Because the cooling time is very long ($t_{\rm cool}\gtrsim$ a few months), the evolution is nearly adiabatic and we expect little energy is radiated.

At later stages, bound debris falls back toward the BH, shocks against itself, and forms an accretion flow, while the unbound ring expands outward.  At the end of the simulation for our fiducial model ($t\simeq 0.6$ days), $t_{\rm cool}$ very near the SMBH is only $\sim 1$~hr, but increases gradually and monotonically outward, reaching a few months in the expanding ring. From $t \sim 10^4$~s onward, the distance at which $t_{\rm cool}=t$ remains constant at $\simeq 70 r_{\rm g}$. 

The fact that the dividing line between regions where $t_{\rm cool} < t$ and where $t_{\rm cool} > t$ stays roughly fixed in place allows us to split the entire system into three regions depending on $t_{\rm cool}/t$: (1) the inner region of the hot accretion flow ($r \lesssim 70 r_{\rm g}$), where $t_{\rm cool}\lesssim t$; (2) the rest of the hot accretion flow, where $t_{\rm cool}\gtrsim t$; and (3) the expanding ring, which has the longest cooling time ($\sim 0.1 - 1$~yr).  This distinction is important because radiation transfer can reach a steady-state only when the photon diffusion time (here essentially $t_{\rm cool}$) is comparable to or shorter than the evolution time.  Put another way, the probability distribution function for the emergence of photons from an optically thick region cuts off much more sharply than linearly for times longer than the photon diffusion time; hence estimating the luminosity by the ratio of thermal energy to cooling time is valid only for $t_{\rm cool} \lesssim t$; when the cooling time is longer, using this ratio leads to a severe overestimate of the luminosity.

 We therefore begin our estimate of the luminosity with region (1), where our methods are most secure. 

\begin{figure}
	\centering
	\includegraphics[height=6.5cm,angle=0]{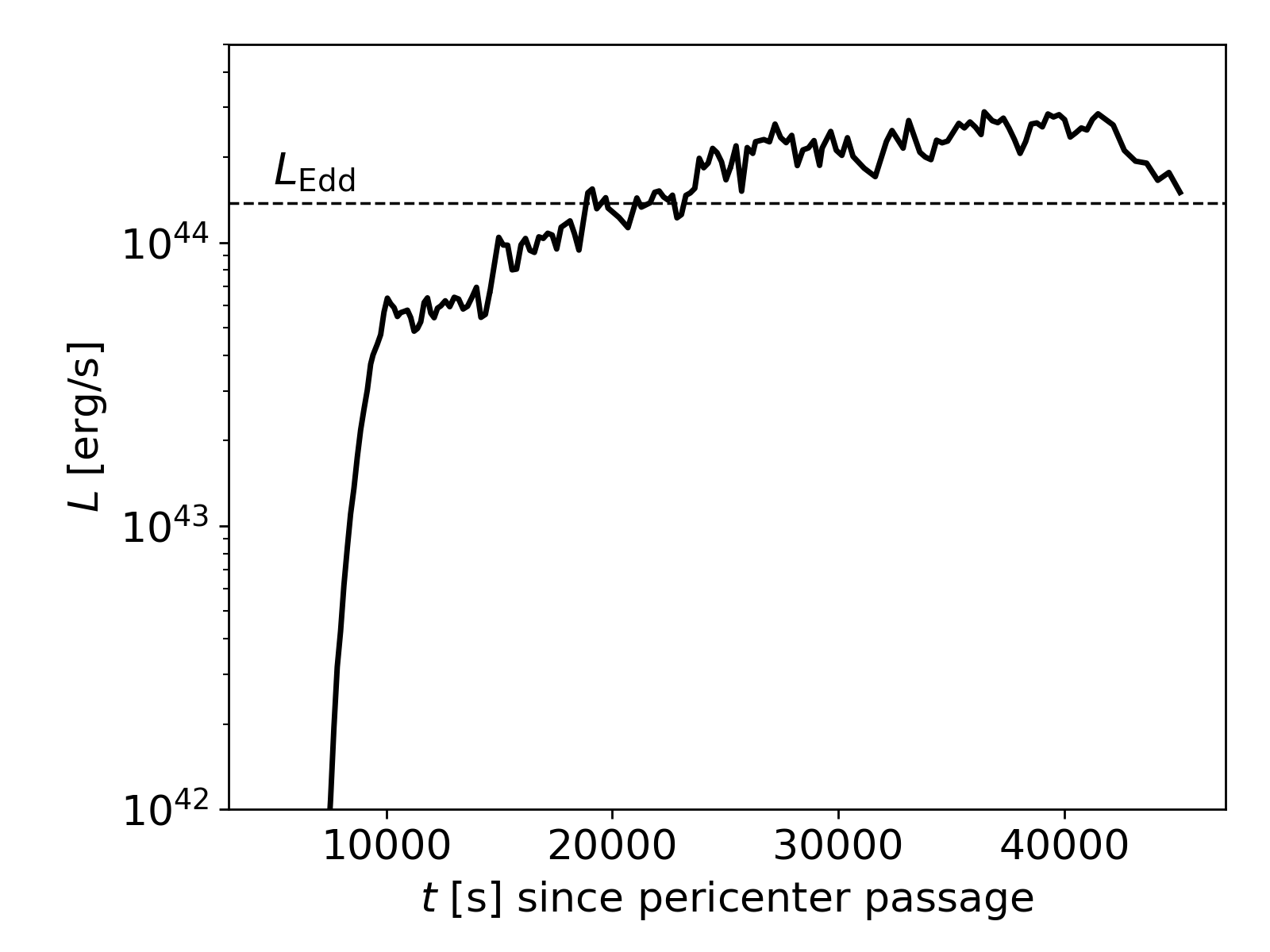}
\includegraphics[height=6.5cm,angle=0]{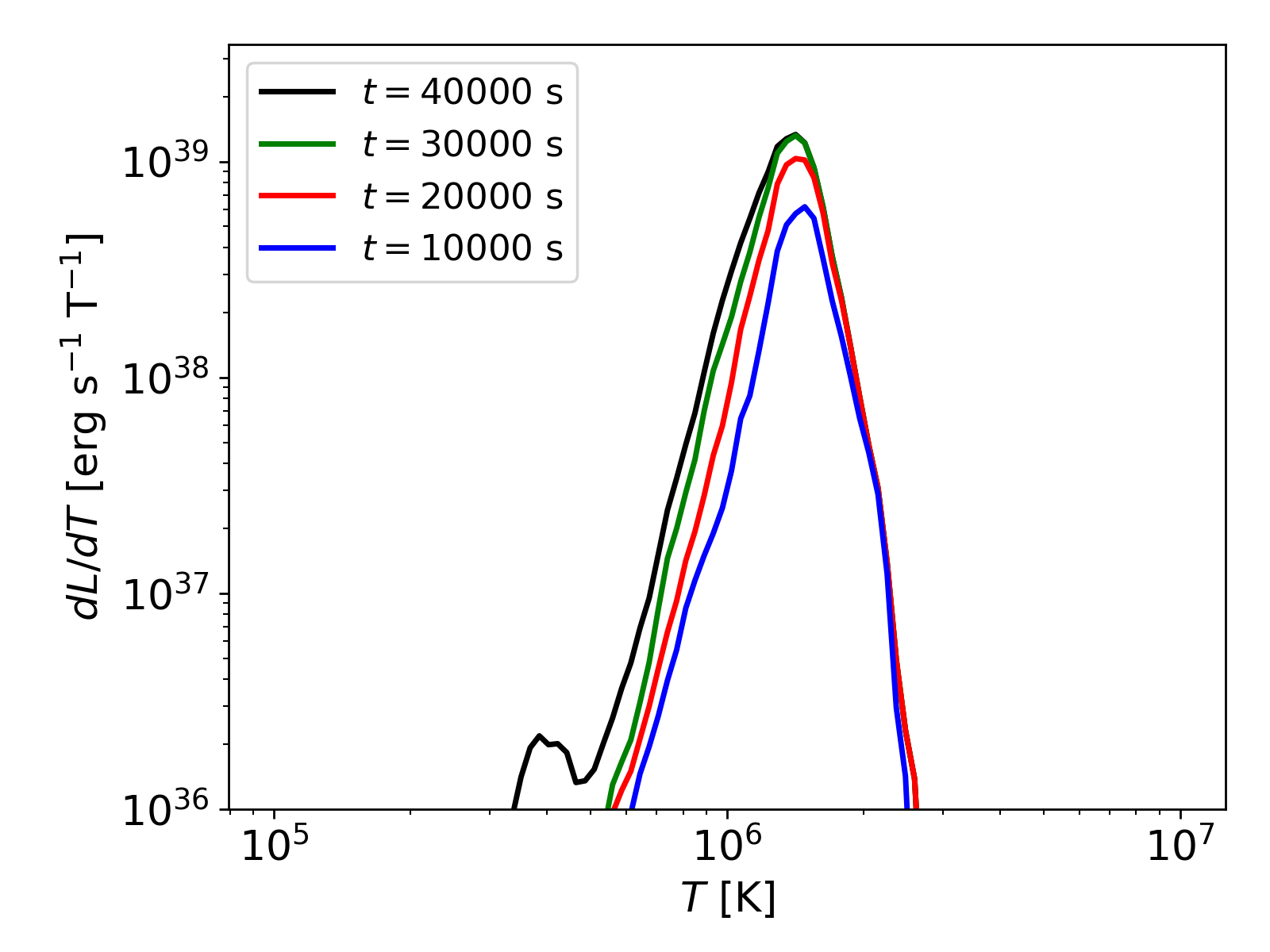} 
\caption{The luminosity estimated using Equation~\ref{eq:Ledd} as a function of time (\textit{left}) and the distribution of the temperature $dL/dT$ at a few different times (\textit{right}). The horizontal dashed line in the \textit{left} panel indicates the Eddington limit for $M_{\rm BH}=10^{6}\Msol$.}\label{fig:luminosity}
\end{figure}

We estimate its total luminosity by integrating the local emissivity of the individual cells with respect to polar angle within the thermalization photosphere at $\sqrt{\tau_{\rm T}\tau_{\rm ff}}\simeq 1$. Here, $\tau_{\rm T}$ ($\tau_{\rm ff}$) is the Thomson (absorption) optical depth integrated inwards from the $\theta$ boundary along the $\theta$-direction. The luminosity from cells in each column above the mid-plane along the polar axis is calculated as,
 \begin{align}
     l_{\rm up}(r, \phi) = \int_{\pi/2}^{\theta_{\rm ph,up}} a T^{4} t_{\rm cool}^{-1}r \sin\theta d\theta,
 \end{align}
 where $a$ is the radiation constant and $\theta_{\rm ph,up}$ is the polar angle of the cell closest to the photosphere for given $r$ and $\phi$ above the mid-plane. $l$ below the mid-plane ($l_{\rm down}$) is calculated similarly by integrating from $\pi/2$ to $\theta_{\rm ph, down}$.  To find the total luminosity, we integrate $l$ for each ($r$, $\phi$) on the grid with $t_{\rm cool}<t$ gives the total luminosity $L$,
 \begin{align}\label{eq:Ledd}
 L = \frac{1}{2}\int_{0}^{2\pi} \int_{r = R_{\rm in}}^{r (t_{\rm cool}<t)} (l_{\rm up}+l_{\rm down}) r dr d\phi,
 \end{align}
 where $R_{\rm in}$ is the radius of the inner radial cutout and $\theta_{\rm ph,up} (\theta_{\rm ph,down})$ is the polar angle of the photosphere above (below) the mid-plane. The effective temperature at each individual cell near the photosphere is then calculated as $T = (l/\sigma)^{1/4}$, where $\sigma$ is the Stefan–Boltzmann constant. We find that after the accretion flow forms ($t \sim 10^4$~s), the total luminosity remain roughly constant in time at $\sim (2-3)\times 10^{44}$~erg~s$^{-1}$, which is roughly the Eddington luminosity for our $10^6 M_\odot$ SMBH, as demonstrated in the \textit{left} panel of Figure~\ref{fig:luminosity}. And the temperature remains at $T\simeq (1-2)\times 10^{6}$ K (see the \textit{right} panel of Figure~\ref{fig:luminosity}).
That its luminosity is roughly Eddington should not be surprising; dimensional analysis alone shows that the Eddington luminosity is the characteristic cooling rate of any plasma whose opacity is close to Thomson and is supported by radiation pressure against gravity \citep{Krolik2010}.

Estimating the luminosity for regions (2) and (3) is 
 more difficult because its radiation transport is not in a steady-state.  The flux reaching the surface may not be well-estimated by $U_{\rm rad}/t_{\rm cool}$; in addition, a diffusion time longer than the dynamical time means that the radiation pressure can do work on the matter, transforming photon energy in gas kinetic energy, or vice versa.  Qualitatively, we might expect that at later times in region (2), the gas is likely to fall inward by a factor of several; the compression should increase its total energy by the same factor.  At the same time, however, its cooling time should increase by the same lengthscale ratio because the gas's scale length, but not its optical depth, changes.  On this basis, we will crudely estimate its contribution to the luminosity from $U_{\rm rad}/t_{\rm cool}$ at the end of the simulation; this yields $\sim 3 \times 10^{43}$~erg~s$^{-1}$.  It might therefore be less luminous than the inner region by a factor of a few.

In region (3), the radiation escape time is a great deal larger than the simulated evolution time.   Because this region moves outward, the radiation energy it carries is reduced by the work done in adiabatic expansion.  How rapidly this occurs can be estimated by examining the time-scaling of this region as revealed by the simulation.  Both the cooling time and the total radiation energy contained within the expanding ring follow simple power-laws, $t_{\rm cool}\propto t^{-0.8}$ and $U_{\rm rad}\propto t^{-0.4}$.  Extrapolating these power-laws out to the time at which $t_{\rm cool}=t$ allows us to predict the luminosity when the photons from this region actually can escape.  At this time ($\simeq 5$~days), we find that the radiation energy has diminished to $\simeq 3.4\times10^{48}$~erg.  This implies a luminosity from the expanding ring of $\simeq 7\times 10^{42}$~erg~s$^{-1}$ during a period of several days around $t \simeq 5$~days. This is rather less than the the emission from region (2) at this stage.

\end{document}